\documentclass[aps,pra,twocolumn,amsmath,amssymb,superscriptaddress,longbibliography]{revtex4}
\usepackage[english]{babel}

\usepackage{amssymb}
\usepackage{dcolumn}
\usepackage{bm}
\usepackage{graphicx}
\usepackage{amsmath}

\usepackage{graphicx}        % standard LaTeX graphics tool
\graphicspath{{pict/}{}}

\usepackage{dcolumn}%%%%%new
\usepackage{bm}
\usepackage[pdfstartview=FitH, CJKbookmarks=true, bookmarksnumbered=true, bookmarksopen=true, colorlinks=true, pdfborder=001, citecolor=blue, linkcolor=blue, urlcolor=blue, linktocpage=true] {hyperref}

\begin{document}
\title{Topological pumping assisted by  Bloch oscillations }
	
\author{Yongguan Ke}
\affiliation{Guangdong Provincial Key Laboratory of Quantum Metrology and Sensing $\&$ School of Physics and Astronomy,Sun Yat-Sen University (Zhuhai Campus), Zhuhai 519082, China}
\affiliation{Nonlinear Physics Centre, Research School of Physics, Australian National University, Canberra ACT 2601, Australia}
	
\author{Shi Hu}
\affiliation{Guangdong Provincial Key Laboratory of Quantum Metrology and Sensing $\&$ School of Physics and Astronomy,Sun Yat-Sen University (Zhuhai Campus), Zhuhai 519082, China}
	
\author{Bo Zhu}
\affiliation{Guangdong Provincial Key Laboratory of Quantum Metrology and Sensing $\&$ School of Physics and Astronomy,Sun Yat-Sen University (Zhuhai Campus), Zhuhai 519082, China}

\author{Jiangbin Gong}
\email{phygj@nus.edu.sg}
\affiliation{Department of Physics, National University of Singapore, Singapore 117551, Republic of Singapore}

\author{Yuri~Kivshar}
\email{yuri.kivshar@anu.edu.au}
\affiliation{Nonlinear Physics Centre, Research School of Physics, Australian National University, Canberra ACT 2601, Australia}
\affiliation{ITMO University, St. Petersburg 197101, Russia}

\author{Chaohong Lee}
\email{lichaoh2@mail.sysu.edu.cn}
\affiliation{Guangdong Provincial Key Laboratory of Quantum Metrology and Sensing $\&$ School of Physics and Astronomy,Sun Yat-Sen University (Zhuhai Campus), Zhuhai 519082, China}
\affiliation{State Key Laboratory of Optoelectronic Materials and Technologies, Sun Yat-Sen University (Guangzhou Campus), Guangzhou 510275, China}

\begin{abstract}
Adiabatic quantum pumping in one-dimensional lattices is extended by adding a tilted potential to probe better topologically nontrivial bands.
This extension leads to almost perfectly quantized pumping for an arbitrary initial state selected in a band of interest, including Bloch states. In this approach, the time variable offers not only a synthetic dimension as in the case of the Thouless pumping, but it assists also in the uniform sampling of all momenta due to the Bloch oscillations induced by the tilt. The quantized drift of Bloch oscillations is determined by a one-dimensional time integral of the Berry curvature, being effectively an integer multiple of the topological Chern number in the Thouless pumping. Our study offers a straightforward approach to yield quantized pumping, and it is useful for probing topological phase transitions.
\end{abstract}
	
\date{\today}
\maketitle

Adiabatic quantum pumping via slow and periodic modulation in certain system parameters has been of tremendous theoretical and experimental interests. It has been investigated with a variety of platforms including, for example,  electrons~\cite{Thouless1983,King1993,Pothier_1992,switkes1999adiabatic,blumenthal2007gigahertz,Xiao2010}, photons~\cite{Kraus2012,ke2016topological,Tangpanitanon,zilberberg2018}, cold atoms~\cite{salger2009directed,Lu2016,lohse2016thouless,Nakajima2016,Ke2017,lohse2018exploring,Hu2019,Lin2020}, and nitrogen-vacancy centers in a diamond \cite{Ma2018}.
Adiabatic pumping connects the underlying geometrical or topological features of a system with its transport behavior.
In practice it is useful in electric current standards~\cite{Pekola2013}, gravimetry~\cite{steffen2012digital,Ke2018}, generation of entangled states~\cite{mandel2003controlled,blanco2018topological,hu2020topological}, and quantum state transfer~\cite{hu2020topological,Mei2018}.
Quantum adiabatic pumping yields both non-quantized transport, such as geometric pumping~\cite{Lu2016}, ratchet transport~\cite{Poletti2008,Salger2013} and edge-state transport~\cite{Kraus2012}, and quantized transport such as Thouless pumping~\cite{Thouless1983,King1993} and its extension in Floquet topological phases~\cite{Ho2012,zhou2014aspects}.
Of particular relevance to this work, the non-quantized geometric pumping in a lattice can be determined by the Berry curvature at a certain momentum value~\cite{Lu2016}, whereas the quantized Thouless pumping yields a topological invariant, namely, the Chern number of a band on a 2-dimensional torus formed by the quasi-momentum and the time variable as a second synthetic dimension.
%Recently, geometric~\cite{Lu2016} and Thouless~\cite{lohse2016thouless,Nakajima2016} pumpings with cold atoms %and generalized Thouless pumping~\cite{Zhou2015,Wang2015,Ma2018} with a single spin were observed in %experiment.

Thouless pumping requires a uniformly filled band (either coherently or incoherently), because its quantization arises as a consequence of equal-weight contributions from the Berry curvatures at all momentum values~\cite{Thouless1983,King1993}.
This feature of Thouless pumping makes it possible to dynamically manifest topological band Chern numbers, which are crucial to understand the integer quantum Hall effect and Chern insulators~\cite{Chiu2016,Bansil2016,haldane2017nobel}.
%This is because all momentum states in the filled band are equally involved to determine global topological properties.
In a fermionic system, the uniform band occupation could be automatically achieved if the band lies below the fermion surface.
For bosons, it becomes highly nontrivial to explore Berry curvatures at all momentum values.
In actual quantum pumping experiments where quantum transport is measured (e.g., via the imaging of a cloud of cold atoms), one resorts to some localized initial states to approximate a Wannier state that uniformly fills a band of interest~\cite{aidelsburger2015,lohse2016thouless,Nakajima2016,lohse2018exploring}.
Furthermore, in probing Floquet topological insulators as non-equilibrium topological matter~\cite{Oka2009, Kitagawa2011,NP2011,Ho2012,cayssol2013floquet,zhou2014aspects,Ho2014,Zhou2018,Loss2016,Eckardt2017}, it is even more involving to experimentally implement the uniform occupation on one particular non-equilibrium quasi-energy band~\cite{Ma2018,Wang2015}.

In this paper, we propose an experimental-friendly adiabatic pumping scheme to yield quantized pumping, without the requirement of uniform band occupation. The obtained pumping in a lattice system is well quantized, regardless of what initial states on a band of interest are prepared.
To our knowledge, this surprising possibility was not known until now.
The central idea is to exploit a tilted lattice, such that the time variable not only offers a synthetic second dimension, but also assists in the sampling of all momentum values uniformly due to the Bloch oscillations~\cite{bloch1929quantenmechanik,Nenciu}.
Thus, complementing previous efforts in using Bloch oscillations to indirectly help to explore band topology~\cite{Atala2013,duca2015aharonov,aidelsburger2015}, we show that Bloch oscillations can actually be a powerful tool to probe the topological Chern number of a band.
%Here we wonder if it is possible to uncover band topology by driving a single momentum state via Bloch oscillations instead of uniform band occupation.
It is also now clear that band topology may induce a quantized drift in Bloch oscillations, an intriguing result not noticed in previous studies of Bloch oscillations versus band structure~\cite{Hartmann2004,Breid_2006,Dias2007,Larson2010,Witthaut2010,Ke2015,Kartashov2016,Kartashov2016a,Liu2019}.

\begin{figure}[htp]
	\center
	\includegraphics[width=\columnwidth]{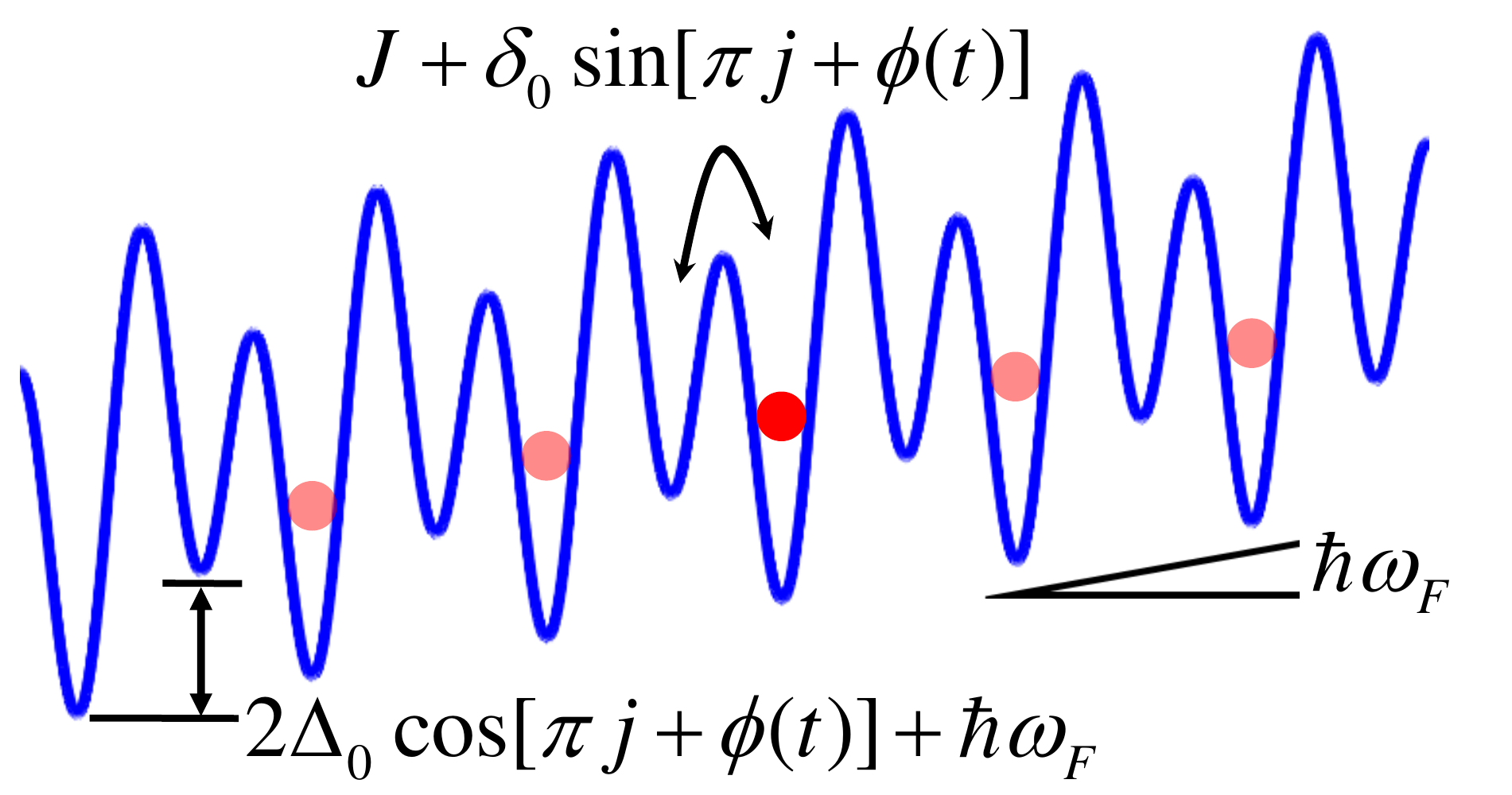}
	\caption{Schematic of adiabatic pumping in a time-modulated superlattice with a tilt. The phase $\phi(t)$ is used to modulate the nearest neighboring hopping and the on-site energy, and $\hbar \omega_F$ is the energy shift associated with the tilt.
	}\label{FigLattice}
\end{figure}

Our adiabatic pumping scheme is depicted in Fig.~\ref{FigLattice}, using a time-modulated super-lattice with a weak tilt.
Certain system parameters are slowly modulated at a frequency $\omega$, $F$ is the energy shift between two neighboring lattice sites due to the tilt, and $\omega_F=F/\hbar$ is half of Bloch oscillation frequency.
For virtually all rational fractions $\omega_{F}/\omega=p/q$, where $p$ and $q$ are co-prime integers, we find that the drift of the system over $q$ modulation cycles is quantized, irrespective of the initial state prepared on a band.
As shown below, such a pumping is determined by  a one-dimensional time integral of the Berry curvature, which  effectively equals to  $q$ times of the Chern number manifested by Thouless pumping.
This is possible due to the effective sampling of all momentum states via Bloch oscillations.
Analogous considerations can be extended to cases with irrational fractions $\omega_F/\omega$, where one recovers a quantized pumping in the long-time average.
The Bloch oscillations themselves carry a new aspect here, because they now can experience a net quantized drift due to the underlying band topology.
%Furthermore, due to the external force, our pumping scheme can maintain the localization of an initial wavepacket, which is in sharp contrast to conventional Thouless pumping with severe delocalization.
%This work hence offers a straightforward and powerful tool for the detection of band topology.}

Without loss of generality, consider a rather simple model adapted from the seminal Rice-Mele model~\cite{Rice1982}: particles are moving in a time-modulated superlattice subjected to an external force, with the following Hamiltonian,
\begin{eqnarray}\label{Ham}
\hat H(t) &=& \sum\limits_j^{} {\left\{\left\{ {J + {\delta _0}\sin [\pi j + \phi(t)]} \right\}\hat a_j^\dag {\hat a_{j + 1}} + h.c.\right\}}\nonumber\\
&& +\sum\limits_j^{} {\left\{ {{\Delta _0}\cos [\pi j + \phi(t) ] + \hbar \omega_F j} \right\}\hat n_j}. \label{Ham}
\end{eqnarray}
Here, $\hat a_j^\dag$ creates a boson  at the $j$-th site and $\hat n_j=\hat a_j^\dag \hat a_j$ is the density operator.
$J$ is the hopping constant.
$\delta_0$ and $\Delta_0$ are the amplitudes of modulations in the hopping strength and the onsite energy, respectively.
$\hbar \omega_F$ is due to a tilt, which can be realized by applying a magnetic field gradient or aligning the superlattice along the gravity.
For convenience, we set $\hbar=1$ by default hereafter.
If $F$ is absent, the model reduces to the Rice-Mele model~\cite{Rice1982}.
The bipartite superlattice can be created by superimposing a simple standing-wave laser with a second double-frequency one.
The phase modulation $\phi(t)=\phi_0+\omega t$ can be realized by tuning the relative phase between two standing-wave lasers and thus the modulation period is given by $T_m=2\pi/\omega$. % with the modulation frequency $\omega$.

In the absence of a tilt,   the Hamiltonian in momentum space is given by $\hat H(k,t)=h_x \hat \sigma_x+h_y\hat \sigma_y+h_z\hat \sigma_z$,
where the effective magnetic field $(h_x,h_y,h_z)=\{2J\cos (k),2{\delta _0}\sin [\phi (t)]\sin(k),{\Delta _0}\cos [\phi (t)]\}$.
By diagonalizing the Hamiltonian, $\hat H(k,t)|u^{0}(k,t)\rangle=\varepsilon^{0} (k,t)|u^{0}(k,t)\rangle$, we analytically obtain the instantaneous eigenvalues $\varepsilon_{\pm}^{0}(k,t)=\pm\sqrt{h_x^2+h_y^2+h_z^2}$ and the corresponding eigenstates $|u_{\pm}^{0}(k,t)\rangle$ (see Appendix~\ref{MomentumSpace}).  In the presence of a tilt, we can obtain analogous solutions by
making a unitary transformation $a_j^\dag=e^{-i\omega_Fjt}b_j^\dag$.  The Hamiltonian then becomes $\hat H_{\rm rot}(t) = \hat H_{1} + \hat H_{2}$ with $\hat H_{1} =\sum_j^{} {\left\{ \left\{ {J + {\delta _0}\sin [\pi j + \phi (t)]} \right\}{e^{i\omega_Ft}}\hat b_j^\dag {\hat b_{j + 1}} + h.c.\right\}}$ and $\hat H_{2}=\sum_j^{} {\left\{ {{\Delta _0}\cos [\pi j + \phi (t)]} \right\}\hat n_j}$.  
That is, the tilt is equivalent to adding a time-dependent phase factor to the hopping term.
This also means that we have changed the boundary condition of the original lattice under PBC. 
However, so long as the dynamic is far way from the boundary, this change of boundary condition has no physical effects.  
As such, all the instantaneous bands and eigenstates can be found by replacing $k$ by $k-\omega_Ft$, with the modified eigenstates $|u_{\pm}(k,t)\rangle =|u_{\pm}^{0}(k-\omega_F,t)\rangle$ and modified dispersion relation   $\varepsilon_{\pm}(k,t)= \varepsilon_{\pm}^{0}(k-\omega_Ft,t)$.
%Under periodic boundary conditions, we obtain the Hamiltonian in quasi-momentum space $H_{rot}(t)=\sum_k H(k,t)$ with
%$\hat H(k,t)=h_x \hat \sigma_x+h_y\hat \sigma_y+h_z\hat \sigma_z$,
%where the effective magnetic field $(h_x,h_y,h_z)=\{2J\cos (k-\omega_Ft),2{\delta _0}\sin [\phi (t)]\sin(k-\omega_Ft),{\Delta _0}\cos [\phi (t)]\}$.
%By diagonalizing the Hamiltonian, $\hat H(k,t)|u(k,t)\rangle=\varepsilon (k,t)|u(k,t)\rangle$, we analytically obtain the instantaneous eigenvalues $\varepsilon_{\pm}(k,t)=\pm\sqrt{h_x^2+h_y^2+h_z^2}$ and the corresponding eigenstates $|u_{\pm}(k,t)\rangle$ (see Supplemental Material for more details~\cite{Suppl}). Note that both $\varepsilon_{\pm}(k,t)$ and $|u_{\pm}(k,t)\rangle$ are time-dependent through both the Bloch oscillation frequency $\omega_F$ and the modulation frequency $\omega$.

According to the theorem of adiabatic transport, the group velocity for momentum $k$ in the $n$-th band is contributed by two terms, the energy dispersion and the Berry curvature~\cite{Xiao2010},
\begin{eqnarray}
v_g(k,t)=\frac{\partial \varepsilon_n(k,t)}{\hbar \partial k}+\mathcal F_n(k,t), \label{Velocity}
\end{eqnarray}
where the Berry curvature is given by,
\begin{equation}
\mathcal F_n(k,t)=-2 \textrm{Im}\big[\sum\limits_{n'\ne n}\frac{\langle u_n|\partial_k \hat H|u_{n'}\rangle\langle u_{n'}|\partial_t \hat H|u_{n}\rangle}{(\varepsilon_n-\varepsilon_{n'})^2}\big],
\end{equation}
with $n=\pm$.
Note that if $\mathcal F_{\pm}^{0}(k,t)$ denotes the analogous Berry curvature  of the gapped Rice-Mele model  ($F=0$),  then $\mathcal F_{\pm}(k,t)=\mathcal F_{\pm}^{0}(k-\omega_Ft,t)$. For later use, the topological Chern number manifested in Thouless pumping is given by
\begin{eqnarray}
C_n=\frac{1}{2\pi} \int_0^{2\pi/d} \int_0^{ T_m} \mathcal F_n^{0} (k, t)\, dk dt, \label{OC}
\end{eqnarray}
where $d$ is the size of each unit cell.
 Because energy bands are periodic in $k$, the first term of group velocity $v_g(k,t)$ oscillates with time.
In a static system, the second term vanishes such that the Zak phase plays no role in Bloch oscillations ~\cite{Atala2013}.
However, due to the periodic modulation via $\phi(t)$ and the adiabatic following of the instantaneous eigenstates,  the anomalous velocity due to Berry curvature becomes crucial here.
To that end one first explicitly obtains the associated Berry curvature at time $t$ for the two bands, i.e.,
\begin{equation}
\mathcal F_{\pm}(k,t)=  2J{\delta _0}\omega {\Delta _0}\frac{1-{\cos^2 {{[\phi(t)]}}\cos^2 {{(k - \omega_Ft)}} }}{{{[\varepsilon_{\pm}^{0}(k-\omega_Ft, t)]^3}}}. \label{AnoVelocity}
\end{equation}
%For later use, we also denote $\mathcal F_{\pm}^{0}(k,t)$ as the Berry curvature without a tilt ($F=0)$.  A two-%dimensional integral of $\mathcal F_{\pm}^{0}(k,t)$ over $t$ and $k$ would give the topological Chern number %detectable by the conventional Thouless pumping.

\begin{figure}[!htp]
	\center
	\includegraphics[width=\columnwidth]{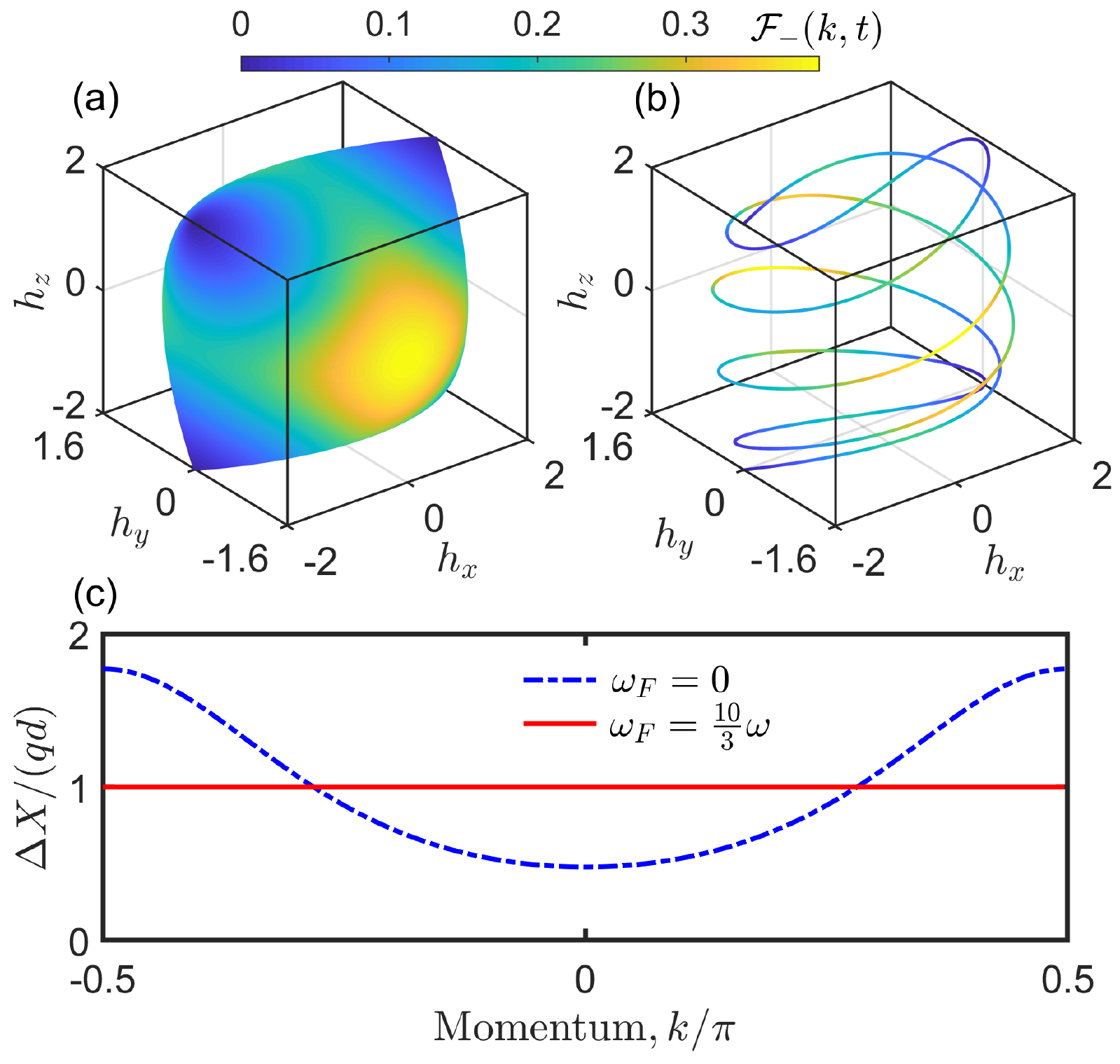}
	\caption{(a) Berry curvature of the lower band, whose magnitudes are represented by color, is indicated on the plot of
	$(h_x,h_y,h_z)$ mapped from $(k,t)$, with $(k,t)$ is made to cover the whole Brillouin zone.
	(b) Same as in (a), but now only $(h_x,h_y,h_z)$ and Berry curvatures at $(k-\omega_F t, t)$ are plotted together.
	(c) The drift of an initial Bloch state over $q$ modulation cycles versus $k$ for $\omega_F/\omega=10/3$ (the red solid line) and $\omega_F=0$ (the blue dashed-dot line). Other parameters are chosen as $J=-1$, $\Delta_0=2$, and $\phi_0=0$.}\label{FigCurvature}
\end{figure}

Consider then a Bloch state as the initial state under our pumping scheme.  To simplify the matters, let us  assume  $\omega_F/\omega=p/q$ as mentioned above, resulting in an overall period $T_{\rm tot}=q T_m$ ($T_m$ is the modulation period in $\phi$).
The amount of pumping at time $\tau$ is simply given by the following semi-classical expression~\cite{Xiao2010},
\begin{equation}
\Delta X(\tau)=\int_0^{\tau} v_g(k, t)dt. \label{SemiClass}
\end{equation}
This expression can be also viewed as the time integral of the quantum flux determined by the group velocity $v_g$.
Because the instantaneous  energy eigenvalues are periodic functions of time and momentum, in an overall period $T_{\rm tot}$ the integral of the dispersion velocity is exactly zero.
Thus only the anomalous velocity due to the Berry curvature can contribute to pumping.  One can find
the drift over the duration of $T_{\rm tot}$ measured by the size of a unit cell $d$,
\begin{equation}
C_{n, \rm {red}}\equiv \frac{\Delta X(q T_m)}{d}=\frac{1}{d}\int_0^{q T_m} \mathcal F_n (k, t)\, dt. \label{SemiClassQuan}
\end{equation}

Our key observation is that $C_{n,\rm red}$ is almost perfectly quantized as a one-dimensional time integral of the Berry curvature $F_n (k, t) =F_n^{0} (k-\omega_Ft, t)$.
Due to the Bloch oscillations at a constant frequency $\omega_F$, all momentum values are uniformly scanned or sampled in this time integral.
Anticipating an effectively ``ergodic'' behavior in such momentum sampling, this integral is hence expected to be independent of the starting value of $\phi$ or equivalently, independent of the initial value of $k$.
This physical intuition is perhaps natural for large integers $q$ and $p$ because highly dissimilar frequencies of $\omega$ and $\omega_F$ enhances the uniformity of the sampling.
Nevertheless, as our results below show, this $k$-independence of the quantum pumping is practically true even when the two frequencies are on low-order resonances.
This being the case, we have
\begin{eqnarray}
C_{n,{\rm red}} &\approx &  \frac{1}{2\pi}  \int_0^{2\pi/d}  \int_0^{q T_m} \mathcal F_n^{0} (k-\omega_Ft, t)\, dk dt,  \ \nonumber \\
&= &   \frac{q}{2\pi} \int_0^{2\pi/d} \int_0^{ T_m} \mathcal F_n^{0} (k, t)\, dk dt = q C_n. \label{effC}
\end{eqnarray}
Thus, $C_{n,\rm{red}}$ is effectively quantized, because it is always very close to $q$ times of the topological Chern number in Thouless pumping.
More discussions on the near perfect quantization of $C_{n, \rm{red}}$ is presented in Appendix~\ref{Relation}.   Clearly then,  compared to the Chern number expression Eq~\eqref{OC} (an integral over a two-dimensional area), $C_{n,\rm{red}}$ defined here as a one-dimensional integral can be regarded as, effectively, a {reduced} expression for the Chern number $C_n$ (apart from the factor $q$).

%In contrast to the Thouless pumping, the upper limit of integral over time is the overall period $q T_m$ instead of the modulation period $T_m$.
%For sufficiently large $\omega_F/\omega$, the transformation $k\rightarrow k+\Delta k$ leads to $\mathcal F(k,t)\rightarrow\mathcal F(k+\Delta k,t+\Delta k/\omega_F)$.
%As $\mathcal F(k,t)$ is a time-periodic function with period $qT_m$, we have $C_{red}^{(n)}(k)=C_{red}^{(n)}(k+%\Delta k)$ for any $\Delta k$, that is, $C_{red}^{(n)}(k)$ is independent of the crystal momentum $k$: $C_{red}^{(n)}(k)=C_{red}^{(n)}$.
%Compared with the conventional Chern number defined in the two-dimensional Brillouin zone,
%\begin{equation}\label{Chern}
%C_{con}^{(n)}=\frac{1}{2\pi}\int_{-\pi/d}^{\pi/d}dk\int_{0}^{T_m}dt \mathcal F_n(k,t),
%\end{equation}
%we can prove that $C_{red}^{(n)}$ is equal to $q$ times of the conventional Chern number, $C_{red}^{(n)}= q %C_{con}^{(n)}$, see Supplemental Material for more details~\cite{Suppl}.

If $C_{n}$ is nonzero, there must be a Dirac monopole at the band-crossing point $h_x=h_y=h_z=0$.
The Berry curvature represents a fictitious magnetic field due to the Dirac monopole and $C_n$ can be viewed as the magnetic flux of the Dirac monopole (up to a $4\pi$ factor) in the ($h_x, h_y, h_z$) parameter space.
For a uniformly filled band, the adiabatic pumping in $\phi(t)$ and the uniform band occupation ensure that an entire surface enclosing the Dirac monopole is fully covered [see Fig.~\ref{FigCurvature}(a)].
In our scheme with a tilt and a Bloch state as the initial state, the sampled $(h_x, h_y, h_z)$ and the associated Berry curvatures rotate around the Dirac monopole due to time evolution itself in a common time period of $qT_m$ [see Fig.~\ref{FigCurvature}(b)].
Somewhat analogous to Ampere's law where a current yields a winding magnetic field, here a rotating field induces pumping and hence a current.

To confirm our insights above, Fig.~\ref{FigCurvature}(c) depicts the drift over $T_{\rm tot}=qT_m$ as a function of $k$, the initial value of a Bloch state, for  $J=-1$, $\Delta_0=2$, $\phi_0=0$ and $\delta_0=0.8$.   The solid line in Fig.~\ref{FigCurvature}(c) is for $\omega_F/\omega=10/3$,  where deviation of $ C_{n, {\rm red}}$  from quantization is not detectable.  That is, the ratio of $\Delta X(q T_m)$  to $qd$ is indeed extremely close to unity, hence  $C_{n, {\rm red}}$ is extremely close to $q$, for any value of $k$, consistent with $C_n=1$.
This is in sharp contrast to the case with $F=0$ [dashed line in Fig.~\ref{FigCurvature}(c)], where the geometrical pumping is not quantized and strongly depends on $k$.   In Appendix~\ref{Relation},  we have also investigated many other cases with different rational ratios of  $\omega_F/\omega$.  Even for very-low-order resonances (e.g., $q=2$), the deviation of the pumping from quantization is about one percent only, a precision that is more than sufficient for  $C_{n, \rm{red}}$ to serve as an effective topological invariant to detect topological phase transitions.
%The

It is also interesting to discuss the cases with irrational $\omega_F/\omega$.   The system's group velocity $v_g$ is then quasi-periodic in time.
In essence that represents cases with $q$ approaching infinity. Hence the pumping is not expected to be well quantized for a duration $NT_m$ with a small $N$.
Nevertheless, the averaging pumping over a sufficiently long time, i.e.,
$\frac{\Delta X(N T_m)}{Nd}\big|_{ N\rightarrow \infty}$
is still quantized because of two reasons. First,  the Berry curvature part of $v_g$ can now sample all momentum values in a more ergodic fashion.  Second,  the time integral of the quasi-periodic dispersion velocity over a sufficiently long time vanishes.   More details can be found in  Appendix~\ref{IrrationalGuassian}.

Given that even an arbitrary Bloch state yields essentially quantized pumping,  it becomes obvious that quantized pumping survives for any initial wavepacket prepared on a band of interest.  To demonstrate this we consider an initial Gaussian wavepacket localized at momentum $k_0=0$ of the lower energy band.  The initial wavefunction at the $j$-th site is hence given by
\begin{equation}
\psi_j(0)=\mathcal{N} e^{-\frac{(j-j_0)^2}{4\sigma^2}} u_{-,j} (k_0,0) e^{i k_0 j},
\end{equation}
Here, $\mathcal{N}$ is a normalization factor, $\sigma$ is the initial wavepacket width, $u_{-,j} (k_0,0)$ represents the instantaneous lower-band spinor eigenstate in the sublattice degree of freedom at time zero.  As the bias between the two sublattices increases, this state dominantly occupies the odd lattice sites. Such type of wavepacket can be prepared by applying an additional harmonic trap~\cite{Lu2016}.
We then examine the density distribution profile $|\psi_j(t)|^2$ of the time-evolving wavepacket $\psi_j(t)$ and the mean displacement
\begin{equation}
\Delta X(t)=X(t)-X(0),
\end{equation}
where $X(t)=\sum_j j|\psi_j(t)|^2$.

%For wider wavepacket, the less momentum states are occupied and hence wavepacket has longer-time coherence (see Supplemental Material for more details~\cite{Suppl}).
%$u_{-,j}(k_0,0)=\langle k_0,o|u_{-,j}(k_0,0)\rangle$ for $j\in \textrm{odd number}$ and $u_{-,j}(k_0,0)=\langle k_0,e|u_{-,j}(k_0,0)%%\rangle$ for $j\in \textrm{even number}$.
%This kind of wavepacket has been prepared by applying an additional harmonic trap~\cite{Lu2016}.

\begin{figure}[!htp]
	\center
	\includegraphics[width=\columnwidth]{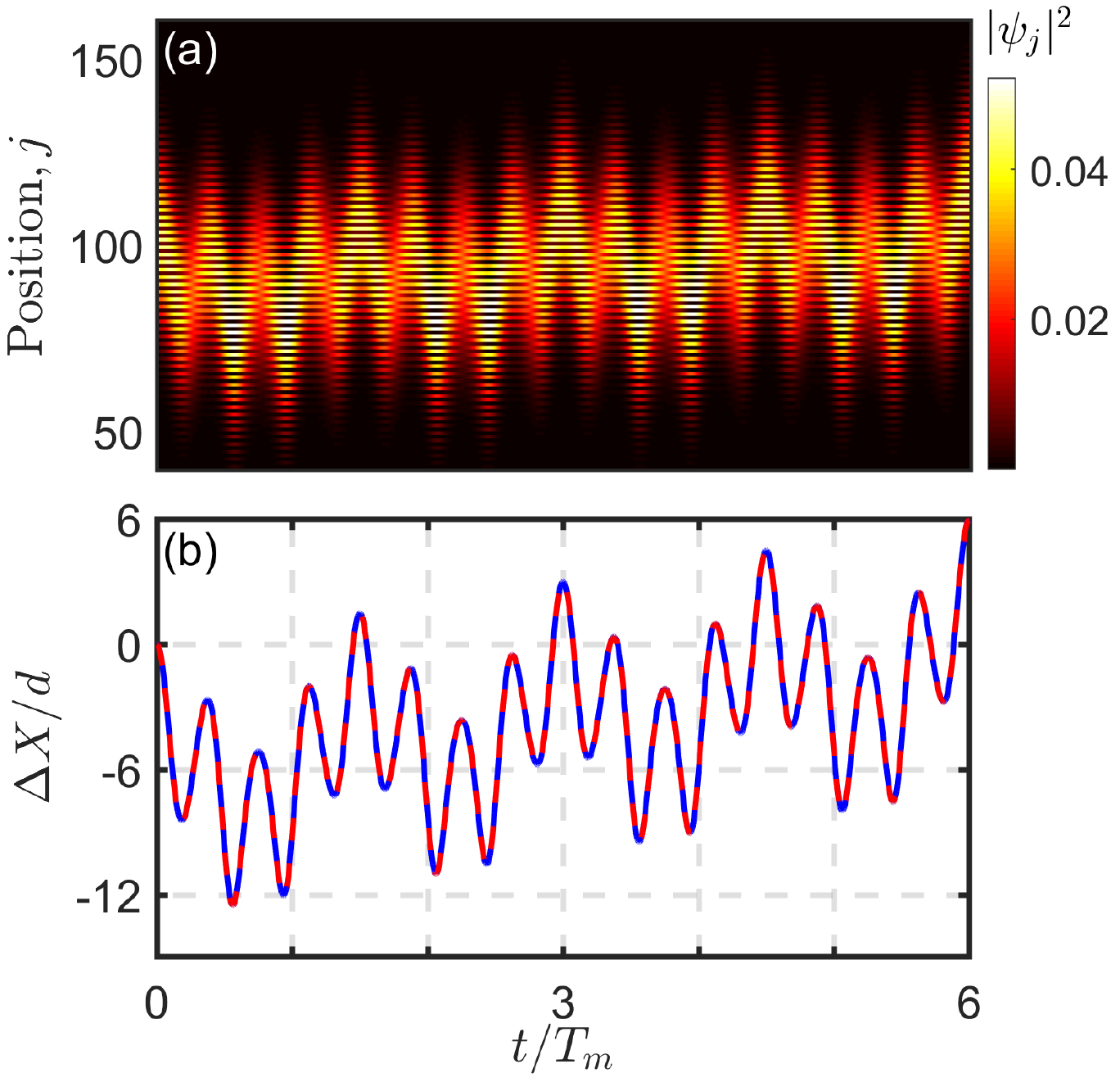}
	\caption{Quantized drifting Bloch oscillations for an initial Gaussian wavepacket in two overall periods.
		(a) Density evolution in real space.
		(b) Displacement as a function of time. The blue solid line and red dashed line are obtained from quantum dynamics calculations and the semi-classical expression Eq.~\eqref{SemiClass}, respectively.
		The parameters are chosen as $J=-1$, $\Delta_0=2$, $\delta_0=0.8$, $\omega=0.03$, $\phi_0=0$, $\omega_F/\omega=4/3$, $d=2$, $\sigma=15$, $j_0=101$ and $k_0=0$.}\label{FigGuassian}
\end{figure}

Fig.~\ref{FigGuassian}(a) and (b) show the time-evolution of wavefunction profile in real space and the drift of the wavepacket center as a function of time.
The instantaneous state can be obtained by iteratively performing the calculations,  $|\psi(t+dt)\rangle=\exp[-i\hat H(t)dt/\hbar]|\psi(t)\rangle$,
with the parameters $J=-1$, $\Delta_0=2$, $\delta_0=0.8$, $\phi_0=0$, $\omega=0.03$, $\omega_F/\omega=4/3$, $\sigma=15$, $j_0=101$.  The frequencies $\omega$ and $\omega_F$ are chosen to be small to ensure adiabatic following, otherwise the quantized pumping breaks down due to the nonadiabatic effects (see Appendix~\ref{Break} for more details).
It is seen that the spatial density profile exhibits cosine-like oscillations with additional modulation, which manifest the Bloch oscillations in a time-modulated system.
More importantly, a quantized drift of such oscillations is seen at multiples of $3T_m$, as displayed by the blue solid line in Fig.~\ref{FigGuassian}(b).
The red dashed line obtained by the semi-classical expression in Eq.~\eqref{SemiClass} perfectly agrees with the one directly obtained by wavepacket dynamics calculations.
Though not shown here,  we have also checked that in the momentum space, the average momentum of the time-evolving state
indeed linearly sweeps the Brillouin zone according to $k=k_0-\omega_Ft$; see Appendix~\ref{EvolutionMomentum}.

\begin{figure}[!t]
	\center
	\includegraphics[width=0.48 \textwidth]{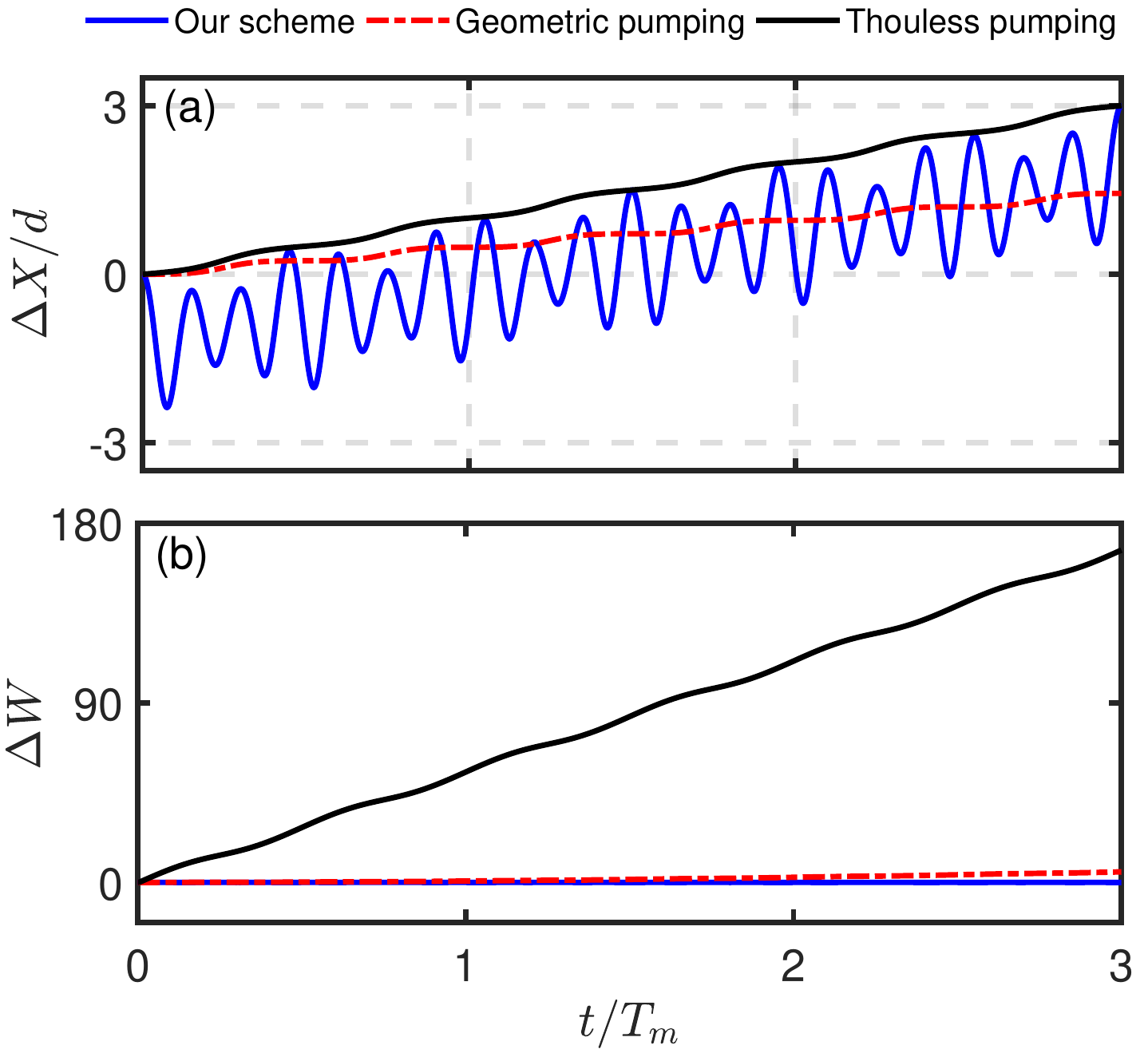}
	\caption{Comparison between topological pumping with a tilt ($\omega_F/\omega=10/3$), geometric pumping with the same initial state but without a tilt, and Thouless pumping with a Wannier state.
	(a) Drift $\Delta X$ versus time $t$.
	(b) Change in the width of time-evolving wavepackets, $\Delta W$ versus time $t$.
	The initial Gaussian wavepacket is parameterized by $\sigma=15$, $j_0=101$ and $k_0=0$.
	The initial state for Thouless pumping is a Wannier state.
	Other common parameters are chosen as $\Delta_0=2$, $J=-1$, $\delta_0=0.8$, $\omega=0.05$, and $\phi_0=0$.
	}\label{FigComparison}
\end{figure}

It is necessary to compare three pumping schemes in a modulated lattice:  (i) topological pumping with a tilt, with the initial state being a wavepacket, (ii) the geometric pumping with the same initial state but without a tilt~\cite{Lu2016}, and (iii)  Thouless pumping where the initial state is a Wannier state~\cite{lohse2016thouless,Nakajima2016}.
In Fig.~\ref{FigComparison}, we show $\Delta X(t)$ and the change in the wavepacket width
\begin{equation}
\Delta W(t)=W(t)-W(0),
\end{equation}
where the wavepacket width is defined as $W (t)= \sqrt{ \sum_j [j-X(t)]^2|\psi_j(t)|^2}$.
In geometric pumping, the transport is not quantized and the wavepacket has insignificant spreading.
In Thouless pumping, although the transport is quantized, the wavepacket exhibits serious spreading even at early time during a pumping cycle.
%Introducing an external force, such diffusion may be partially suppressed while the quantized value maintains the same (see Supplemental Material for details~\cite{Suppl}).
In our topological pumping with a tilt, not only the transport is quantized, but also the wavepacket maintains its spatial localization or coherence over a long time; see Appendix~\ref{EvolutionGuassian}, a feature of considerable interest for quantum state transfer~\cite{Mei2018}.
For completeness, in Appendix~\ref{PumpingWannier} we also show the pumping dynamics of Wannier states in a titled field.
To summarize,  topological pumping with a tilt has advantageous aspects from both geometrical pumping and Thouless pumping.
%yields quantized transport and maintains wavepacket localization and (ii)

In summary, we have put forward a simple scheme of adiabatic pumping by introducing a small tilt to a lattice on top of other time modulation to system parameters.
Quantized pumping can now be readily realized in experiments because it works for arbitrary initial state prepared on a band of interest (see Appendix~\ref{Experiment} for more details).
As such, there is no longer a need to engineer uniform band occupation as in Thouless pumping.
It should be stressed that the resultant pumping is not {\it mathematically} quantized, but in practice it is well quantized with remarkable precision, hence highly useful for probing topological phase transitions.
%We introduce a reduced Chern number, a one-dimensional line integral over time but not the integral over conventional two-dimensional %momentum-time space, which successfully explains the quantized displacements in Bloch oscillations.
%As the external force drives the instantaneous quasi-momentum linearly sweeping through the whole Brillouin zone, the reduced Chern %number is independent of the initial momentum and directly gives the conventional Chern number.
%The role of the external force makes the instantaneous states iterate over the curvature surface, effectively the same as uniform band occupation.
%That is why the reduced Chern number is independent of the initial momentum which only changes the starting point of the trajectory.
%According to Laughlin's argument~\cite{Laughlin1981}, in two-dimensional quantum systems the quantization of Hall conductance results %from an effective topological pumping.
%When electron-electron interactions play a crucial role, the Hall conductance is related to a fractional filling factor~\cite{Tsui1982}.
It would be interesting to extend our results to disordered or many-body systems.
% for deeper understanding the fractional quantum Hall effects as well as other topological orders~\cite{Xiaogang2017}.
Indeed, probing topological invariants without uniform band filling can be important in topological systems without conventional band structures, such as in disordered topological insulators and interacting topological insulators~\cite{Titum2015,rachel2018}.   Our scheme may be also extended to probe topological invariants in high-dimensional systems.

Note added: In the review process, we became aware that the topology of stroboscopic Poincaré orbits can lead to unidirectional transport accompanied by Bloch oscillations in a driven inhomogeneous lattice gas~\cite{cao2020transport}.

\begin{acknowledgments}
Y. Ke and S. Hu made equal contributions. This work has been supported by the Key-Area Research and Development Program of GuangDong Province under Grants No. 2019B030330001, the National Natural Science Foundation of China (NNSFC) under Grants [No. 11874434, No.11574405], and the Science and Technology Program of Guangzhou (China) under Grants No. 201904020024. Y.K. is partially supported by the Office of China Postdoctoral Council (Grant No. 20180052), the National Natural Science Foundation of China (Grant No. 11904419), and the Australian Research Council (DP200101168).   J.G. acknowledges support from Singapore NRF Grant No. NRF-NRFI2017-04 (WBS No. R-144-000-378-281).
\end{acknowledgments}

\appendix

%%%%%%%%%%%%%%%%%%%%%%%%%%%%%%%%%%%%%%%%%%
\section{Hamiltonian in momentum space} \label{MomentumSpace}

In the case of a single particle, to obtain the Hamiltonian in momentum space, we make a Fourier transformation,
\begin{eqnarray}\label{Fourier}
{\hat b_{2j}^{\dagger}} &=& \frac{1}{\sqrt{L}}\sum\limits_k^{} {{e^{ik2j}}{\hat b_{k,e}^{\dagger}}}, \nonumber \\
{\hat b_{2j - 1}^{\dagger}} &=& \frac{1}{\sqrt{L}}\sum\limits_k^{} {{e^{ik(2j - 1)}}{\hat b_{k,o}^{\dagger}}}.
\end{eqnarray}
Here, $k$ is the quasi-momentum, and $e$ ($o$) respectively represents an even (odd) site.
$L$ is the total number of unit cells.
When the tilt is absent, $\hbar \omega_F=0$, substituting Eq.~\eqref{Fourier} into Hamiltonian (1) in the main text, we can obtain Hamiltonian in the quasi-momentum space, $H(t)=\sum_k H(k,t)$ with
\begin{eqnarray}
\hat H(k,t)&=&{\big\{ {2J\cos (k) + 2i{\delta _0}\sin [\phi (t)]\sin(k)} \big\}\hat b_{k,o}^\dag {\hat b_{k,e}}}\nonumber\\
&+& h.c.+ {\Delta _0}\cos [\phi (t)]\big(\hat b_{k,e}^\dag {\hat b_{k,e}}- b_{k,o}^\dag {\hat b_{k,o}}\big). \label{HamK0}
\end{eqnarray}
In terms of Pauli matrices describing the sublattice degree of freedom, the Hamiltonian becomes
$\hat H(k,t)=h_x\hat \sigma_x+h_y\hat \sigma_y+h_z\hat \sigma_z$,
where $(h_x,h_y,h_z)=\{2J\cos (k),2{\delta _0}\sin [\phi (t)]\sin(k),{\Delta _0}\cos [\phi (t)]\}$ are the three components of an effective magnetic field.
One then obtains the eigenvalues and eigenstates by diagonalizing the Hamiltonian, $\hat H(k,t)|u^{0}(k,t)\rangle=\varepsilon^0 (k,t)|u^{0}(k,t)\rangle$. The superscript $0$ denotes zero tilt.
The eigenvalues are given by
\begin{eqnarray}
&&\varepsilon_{\pm}^0=\pm\sqrt{h_x^2+h_y^2+h_z^2}\nonumber \\
&&=\pm\sqrt{4J^2\cos ^2(k)+4{\delta _0}^2\sin^2 [\phi (t)]\sin ^2(k)+{\Delta _0}^2\cos^2 [\phi(t)]},\nonumber
\end{eqnarray}
and the eigenstates without normalization are given by
\begin{equation}
{|u_{\pm}^0\rangle} = \left( {\begin{array}{*{20}{c}}
	{\frac{{2J\cos (k) - 2{\delta _0}i\sin [\phi(t)]\sin (k)}}{{{\varepsilon_{\pm}^0} -{\Delta _0}\cos [\phi (t)]}}}\\
	1
	\end{array}} \right).
\end{equation}

In the presence of a tilt, by making a unitary transformation $a_j^\dag=e^{-i\omega_Fjt}b_j^\dag$, the Hamiltonian (1) is transformed to
\begin{eqnarray}
\hat H_{\rm rot}(t) &=&\sum_j^{} \left\{ \left\{ {J + {\delta _0}\sin [\pi j + \phi (t)]} \right\}{e^{i\omega_Ft}}\hat b_j^\dag {\hat b_{j + 1}}
+ h.c.\right\}\nonumber \\
&+&\left\{ {{\Delta _0}\cos [\pi j + \phi (t)]} \right\}\hat n_j.
\end{eqnarray}
Similarly, the Hamiltonian in the momentum space is given by
\begin{eqnarray}
\hat H(k,t)&=&2{\big\{ {J\cos [K(k,t)] + i{\delta _0}\sin [\phi (t)]\sin[K(k,t)]} \big\}\hat b_{k,o}^\dag {\hat b_{k,e}}}\nonumber\\
&+& h.c.+ {\Delta _0}\cos [\phi (t)]\big(\hat b_{k,e}^\dag {\hat b_{k,e}}- b_{k,o}^\dag {\hat b_{k,o}}\big), \nonumber \\ \label{HamK}
\end{eqnarray}
where $K(k,t)=k-\omega_Ft$. Compared Eq.~\ref{HamK} with Eq.~\ref{HamK0}, all the instantaneous energy bands and eigenstates can be found by replacing $k$ by $k-\omega_Ft$, with the modified eigenstates $|u_{\pm}(k,t)\rangle =|u_{\pm}^{0}(k-\omega_F,t)\rangle$ and modified dispersion relation   $\varepsilon_{\pm}(k,t)= \varepsilon_{\pm}^{0}(k-\omega_Ft,t)$.

\section{Relation between $C_{n,\rm{red}}$ and $C_n$} \label{Relation}

Here we first show that the reduced Chern number $C_{n,\rm red}$ defined in the main text as a one-dimensional time integral is independent of the initial momentum value of  a Bloch state and equal to $q$ times of the conventional Chern number, namely, $C_{n,\rm red}= q C_n$, if $\omega_F/\omega=p/q\rightarrow \infty$.   Consider first the
Berry curvatures
\begin{widetext}
\begin{eqnarray}
\mathcal F_{\pm}(k,t)&=&  \frac{\pm 2J{\delta _0}\omega {\Delta _0}\left\{1-{\cos^2 {{[\phi(t)]}}\cos^2 {{[K(k,t)]}} }\right\}}{{{\left\{4J^2\cos ^2[K(k,t)]+4{\delta _0}^2\sin^2 [\phi (t)]\sin ^2[K(k,t)]+{\Delta _0}^2\cos^2 [\phi (t)]\right\}}^{3/2}}} \nonumber \\
&=&\mathcal F_{\pm}^0(k-\omega_F t,t), \label{AnoVelocity}
\end{eqnarray}
where $\mathcal F_{n}^0(k,t)$ denotes the Berry curvature in the absence of a tilt.  When $k$ is shifted to $k+\Delta k$, and $t$ is shifted to $t+\Delta k/\omega_F$, the Berry curvature is given by
\begin{eqnarray}
&&\mathcal F_{\pm}(k+\Delta k,t+\Delta k/\omega_F) \nonumber \\
&=& \frac{ \pm 2J{\delta _0}\omega {\Delta _0}\left\{1-{\cos^2 {{[\phi(t)+\frac{\omega}{\omega_F}\Delta k ]}}\cos^2 {{[K(k,t)]}} }\right\}}{{{\left\{4J^2\cos ^2[K(k,t)]+4{\delta _0}^2\sin^2 [\phi (t)+\frac{\omega}{\omega_F}\Delta k]\sin ^2[K(k,t)]+{\Delta _0}^2\cos^2 [\phi (t)+\frac{\omega}{\omega_F}\Delta k]\right\}}^{3/2}}} \nonumber  \\
&\cong &  \frac{\pm 2J{\delta _0}\omega {\Delta _0}\left\{1-{\cos^2 {{[\phi(t)]}}\cos^2 {{[K(k,t)]}} }\right\}}{{{\left\{4J^2\cos ^2[K(k,t)]+4{\delta _0}^2\sin^2 [\phi (t)]\sin ^2[K(k,t)]+{\Delta _0}^2\cos^2 [\phi (t)]\right\}}^{3/2}}}\nonumber \\
&=&\mathcal F_{\pm}(k,t).
\label{F2}
\end{eqnarray}
\end{widetext}
The approximately equal sign here can be replaced by an exactly equal sign if $\omega_F/\omega\rightarrow \infty$. Actually, even when $\omega_F$ is comparable to $\omega$, this relation still holds with high precision, an important feature that will become clearer later.
Next we note the following rewriting of one-dimensional time integrals:
\begin{eqnarray}
&&\int_{0}^{qT_m} \mathcal F_{\pm}(k+\Delta k,t+\Delta k/\omega_F)dt \nonumber \\
&=&\int_{\Delta k/\omega_F}^{qT_m+\Delta k/\omega_F} \mathcal F_{\pm}(k+\Delta k,t)dt\nonumber \\
&=&\int_{0}^{qT_m} \mathcal F_{\pm}(k+\Delta k,t)dt,
\end{eqnarray}
where the last equal sign is due to the fact that the Berry curvature is a periodic function of time with period $qT_m$.  Comparing this with Eq.~(\ref{F2}), one immediately has
\begin{eqnarray}
C_{n,\rm{red}}&=&\frac{1}{d}\int_{0}^{qT_m} \mathcal F_{\pm}(k,t) dt \nonumber \\
&=&  \frac{1}{d}\int_{0}^{qT_m} \mathcal F_{\pm}(k+\Delta k,t) dt.
\end{eqnarray}
That is,  $C_{n, \rm{eff}}$ as the time intergral of $\mathcal F_{\pm}(k,t)$ is practically independent of $k$, i.e.,
\begin{equation}
C_{n,\rm red}(k+\Delta k) = C_{n,\rm red}(k),
\end{equation}
for any $\Delta k$.
Now if we consider an averaging  over $k$, we immediately have
\begin{eqnarray}
C_{n,\rm red}&\approx \frac{1}{2\pi}\int_0^{qT_m}\int_{-\pi/d}^{\pi/d} \mathcal F_n^0(k-\omega_F t,t) dt dk \nonumber \\
&=\frac{q}{2\pi}\int_0^{T_m}\int_{-\pi/d}^{\pi/d} \mathcal F_n^0(k,t) dt dk=q C_n.
\end{eqnarray}
The second equal sign is because the Chern number $C_n$ as a two-dimensional integral are the same in each pumping cycle  if there is no external force.
%To show this, we alternately use another expression of the Berry curvature,
%\begin{equation}
%\mathcal F_n=i\big(\langle\partial_{k}u_{n}|\partial_{t}u_{n}\rangle -\langle\partial_{t}u_{n}|%\partial_{k}u_{n}\rangle\big).
%\end{equation}
\begin{figure}[htp]
	\center
	\includegraphics[width=0.48 \textwidth]{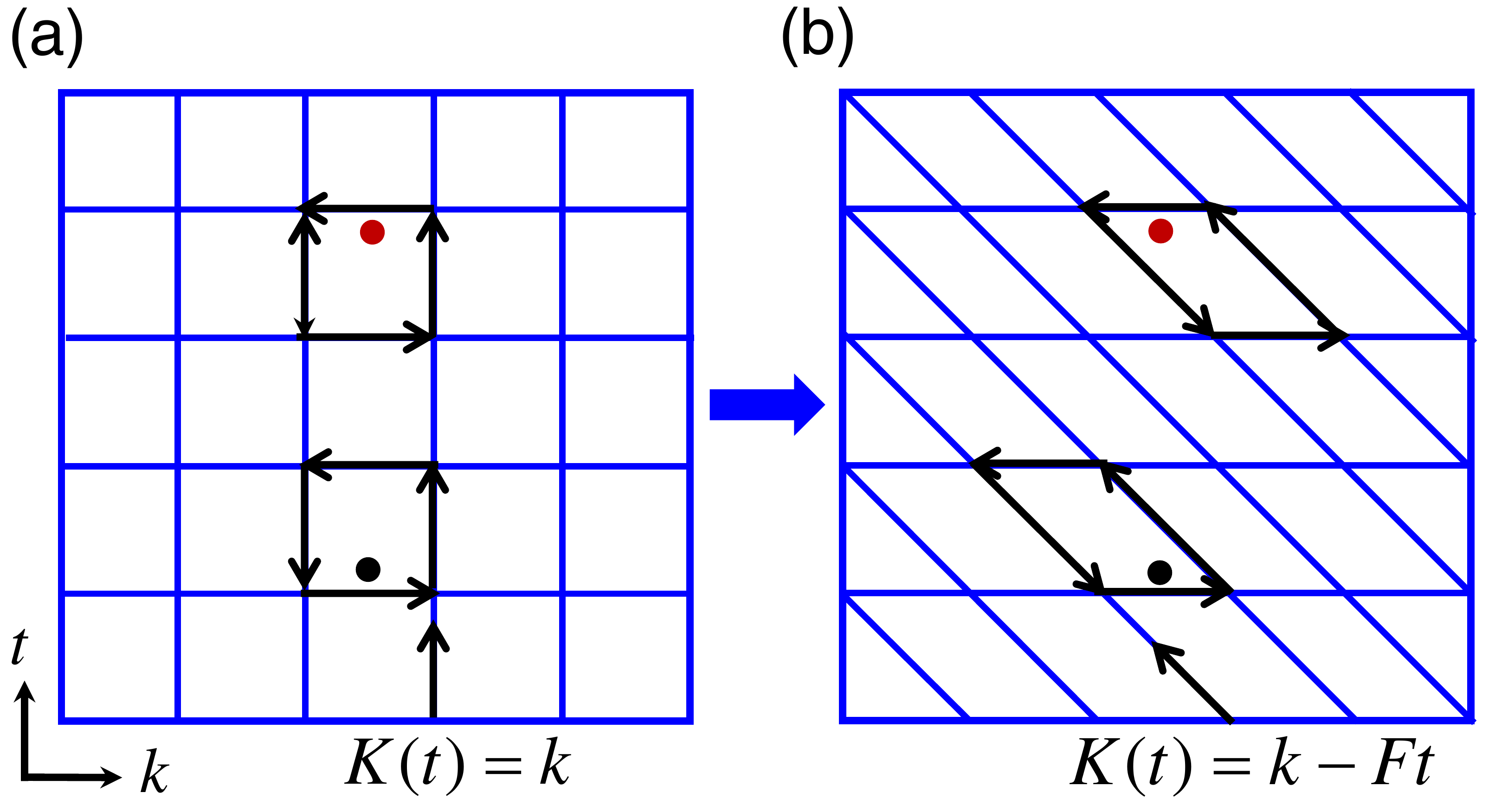}
	\caption{Schematic diagram for numerical calculations of Chern number (a) in the absence of a tilt and (b) inthe presence of a tilt. The red and black dots are the locations of singluarity points at the north and south poles.
	}\label{FigChernNumber}
\end{figure}

In numerical calculations, we discrete the parameter space into mesh grids, and then apply the Stokes theorem to transform the surface integral to a line integral of each grid~\cite{fukui2005chern}.
According to the theory of dynamic winding number, the line integrals encircling singularity points are the major contributions to Chern number~\cite{zhu2019dynamic}.
The Chern numbers for the two bands are the same as those in the original Rice-Mele model, even if we account for the tilt we introduce.
This is because the external force only linearly shift the momentum and  reshape the grids from square to rhombus without affecting the winding number of singularity points; see the schematic diagram in Fig.~\ref{FigChernNumber}.

\begin{figure}[!htp]
	\center
	\includegraphics[width=0.48 \textwidth]{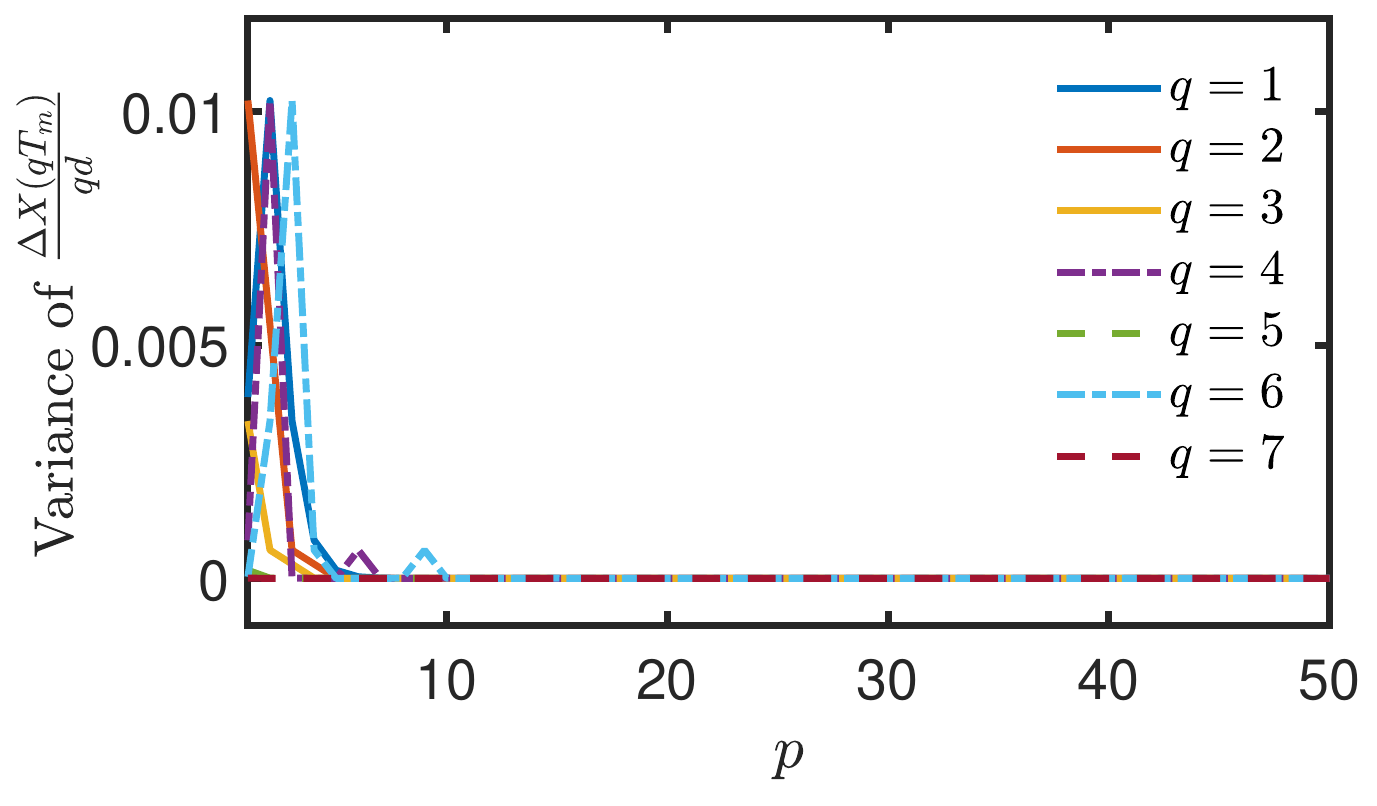}
	\caption{Variance of $\Delta X(qT_m)/(qd)$ as a function of $p=q\omega_F/\omega$. The parameters are chosen as $J=-1$, $\Delta_0=2$, and $\delta_0=0.8$. }\label{FigXForce}
\end{figure}

We calculate standard variance of $\Delta X(q T_m)/(qd )$ over all momentum states as a function of $p=q\omega_F/\omega$ when $q$ ranges from $1$ to $7$.
The result is shown in Fig.~\ref{FigXForce} with parameters $J=-1$, $\Delta_0=2$, $\phi_0=0$, and $\delta_0=0.8$.
The mean displacement per modulation period is exactly $1$ unit cell.
It is clear the deviation of $\Delta X(q T_m)/(qd)$ from $1$ exponentially decays as the ratio between the external force and driving frequency increases.
Taking $q=7$ as an example, the deviation between $\Delta X(7 T_m)/(7d)$ from unity already reaches machine precision/error for $p=11$, i.e. $\omega_F/\omega=11/7$.
This super-fast decay of $\Delta X(N T_m)/(Nd)$ with $N$ also persists even when $N$ is not equal to multiples of $q$.
These numerical results show that the actual condition for the quantization of $C_{n,\rm{red}}$  as an effective topological invariant is much looser than that in the above analysis.

\section{Topological pumping of Gaussian wavepackets} \label{Gaussian}

\subsection{Irrational case} \label{IrrationalGuassian}

\begin{figure}[!t]
	\center
	\includegraphics[width=0.48 \textwidth]{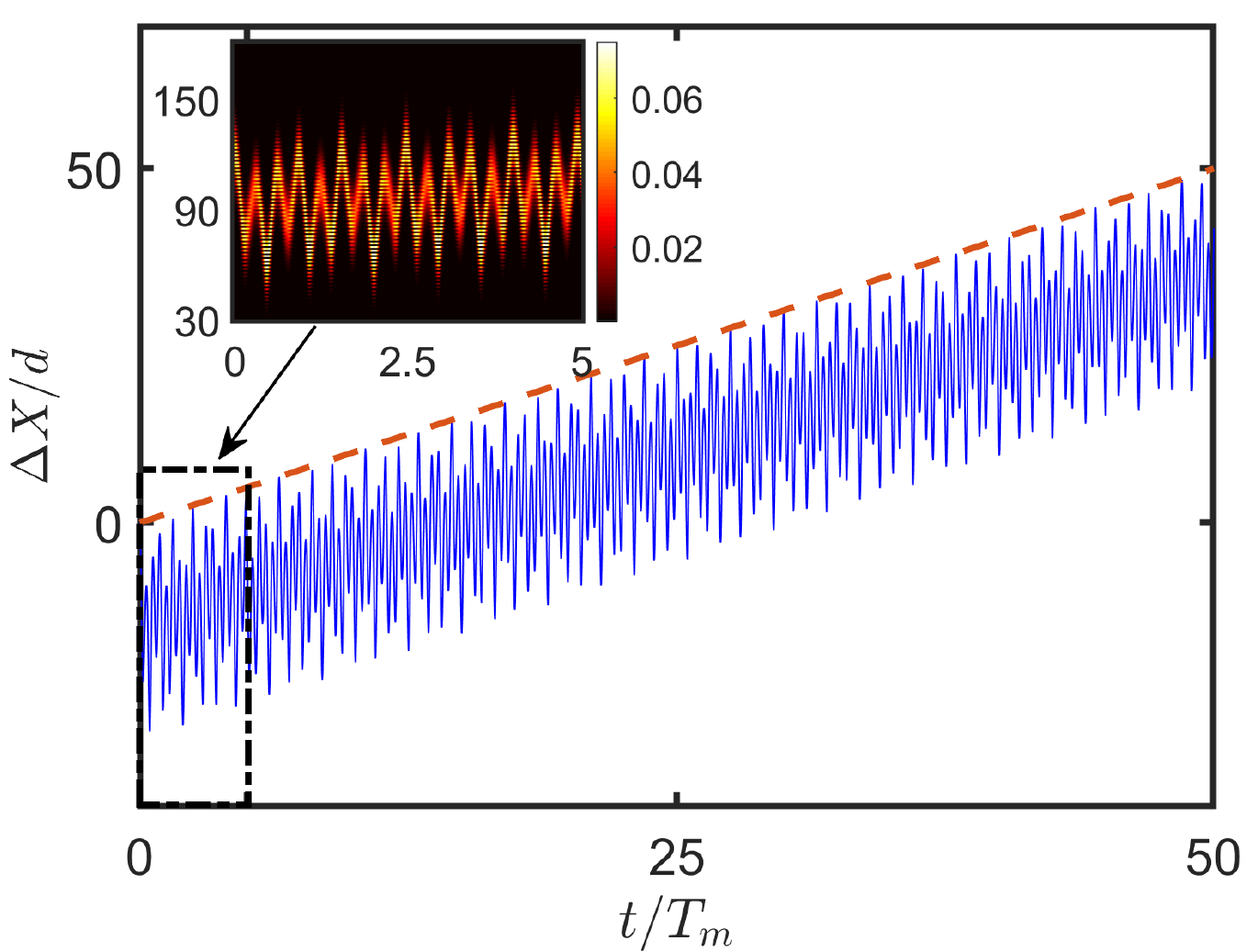}
	\caption{Displacement for an initial Gaussian-like state in an irrational case of $\omega_F/\omega=(\sqrt{5}+1)/2$. The dashed red line is obtained by averaging the velocity in long time.  Inset: Time evolution of density distribution in real space.
		Other parameters are chosen as $J=-1$, $\Delta_0=2$, $\delta_0=0.8$, $\omega=0.01$, $\phi_0=0$, $d=2$, $\sigma=15$, $j_0=120$ and $k_0=0$.}\label{FigLong}
\end{figure}

For the cases where $p$ and $q$ are not commensurate, the eigenvalues are quasi-periodic functions of time and hence the velocity due to the dispersion of bands becomes also quasi-periodic.
This means that the displacement is generally not quantized because the velocity due to the dispersion of bands cannot  self-cancel over a short time.
However, the mean displacement in the long time average,
\begin{equation}
\frac{\Delta X(N T_m)}{ Nd}\big|_{ N\rightarrow \infty}\rightarrow 1,
\end{equation}
because the integral of velocity due to the dispersion of bands over long time  vanishes and only quantized displacement due to anomalous velocity leaves.
For example, we consider $p/q=(\sqrt{5}+1)/2$ and calculate the mean position shift as a function of time via Eq.~(6) in the main text; see Fig.~\ref{FigLong}.
The inset shows the corresponding time-evolution of density distribution in relatively short time.
Other parameters are chosen as $J=-1$, $\Delta_0=2$, $\delta_0=0.8$, $\omega=0.01$, $\phi_0=0$, $d=2$, $\sigma=15$, $j_0=120$ and $k_0=0$.
We can observe quasi-periodic Bloch oscillations accompanied by a linear displacement guided by red dashed line in Fig.~\ref{FigLong}.
After averaging the velocity in infinite time, the red dashed line is given by $\Delta X(NT_m)/d=N$.
The displacement due to anomalous velocity linearly increases with the $N$ multiple of pumping cycle. Because the oscillations have finite width, $W$ (which can be suppressed by strong external force), The fluctuation in displacement $W/\Delta X\approx W/(2N)$ will eventually vanish as the total time under consideration approaches infinity.

\subsection{Rational case: Breakdown of quantized pumping under strong tilting} \label{Break}
\begin{figure}[!t]
	\center
	\includegraphics[width=0.5 \textwidth]{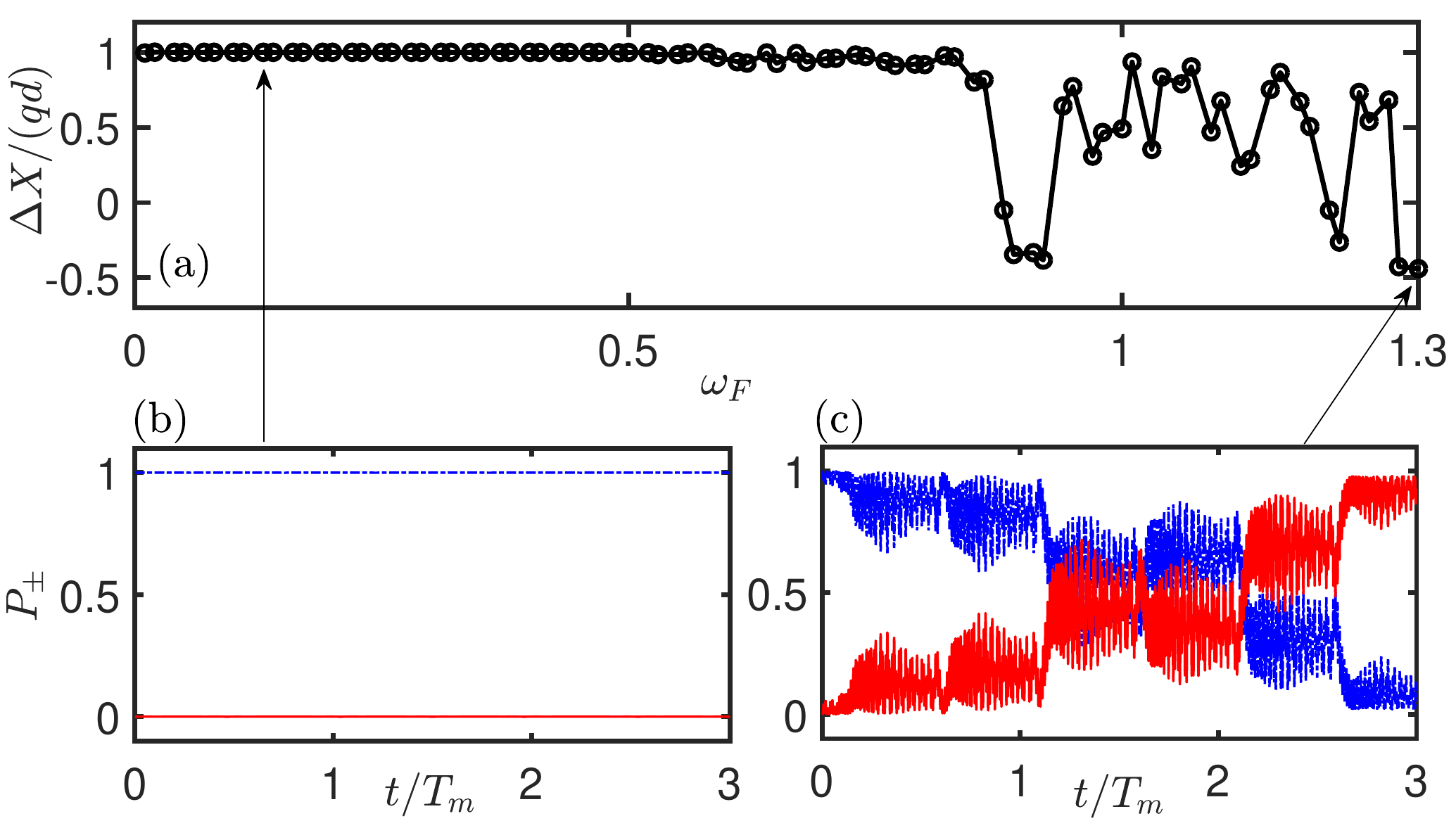}
	\caption{(a) Displacement in an overall period as a function of $\omega_F$. The parameters are chosen as $J=-1$, $\Delta_0=2$, $\delta_0=0.8$, $\omega=0.03$, $\omega_F/\omega=p/3$, $\phi_0=0$, $d=2$, $\sigma=15$, $j_0=120$ and $k_0=0$. (b) and (c): Occupations in the lower band (blue dashed-dot line) and higher band (red solid line) for $\omega_F=0.13$ and $\omega_F=1.3$, respectively. The other parameters are the same as those in (a). }\label{Breakdown}
\end{figure}

We know that larger ratio $\omega_F/\omega$ makes the effective topological invariant $C_{n,\rm{red}}$ closer to the ideal Chern number.
However, in practice Landau-Zener transitions between different bands become so serious for larger $\omega_F$ that the adiabatic condition is no longer satisfied in real time evolution.
We show how the quantized pumping in the lowest band breaks down as $\omega_F$ increases, see Fig.~\ref{Breakdown}(a).
The parameters are chosen as $J=-1$, $\Delta_0=2$, $\delta_0=0.8$, $\omega=0.03$, $\phi_0=0$, $d=2$, $\sigma=15$, $j_0=120$ and $k_0=0$.
$\omega_F/\omega=p/3$ is changed with discrete $p$ (excluding the multiples of $3$).
When $\omega_F$ is modest, the displacement per pumping cycle is close to one unit cell according to the semi-classical expression~\eqref{SemiClass}.
We show the occupations in the lower and higher band as functions of time for modest $\omega_F=0.13$,  see Fig.~\ref{Breakdown}(b).
The Gaussian wavepacket always stays in the lower band and Landau-Zener transition is negligible.
The corresponding displacement per pumping cycle is almost one unit cell, indicated by the pointing arrow.
However, when $\omega_F$ becomes larger (e.g. $\omega_F=1.3$), the wavepacket is rapidly transferred between the lower and the higher bands and vice versa, see Fig.~\ref{Breakdown}(c).
The arrow points to the corresponding displacement in Fig.~\ref{Breakdown}(a).
Since the adiabatic condition is not satisfied, it is impossible to predict the nonquantized displacement via the semi-classical expression~\eqref{SemiClass}.
What is worse, the transport direction is also possibly reversed around some larger $\omega_F$.

\subsection{Density evolution in momentum space} \label{EvolutionMomentum}
\begin{figure}[!htp]
	\center
	\includegraphics[width=0.48 \textwidth]{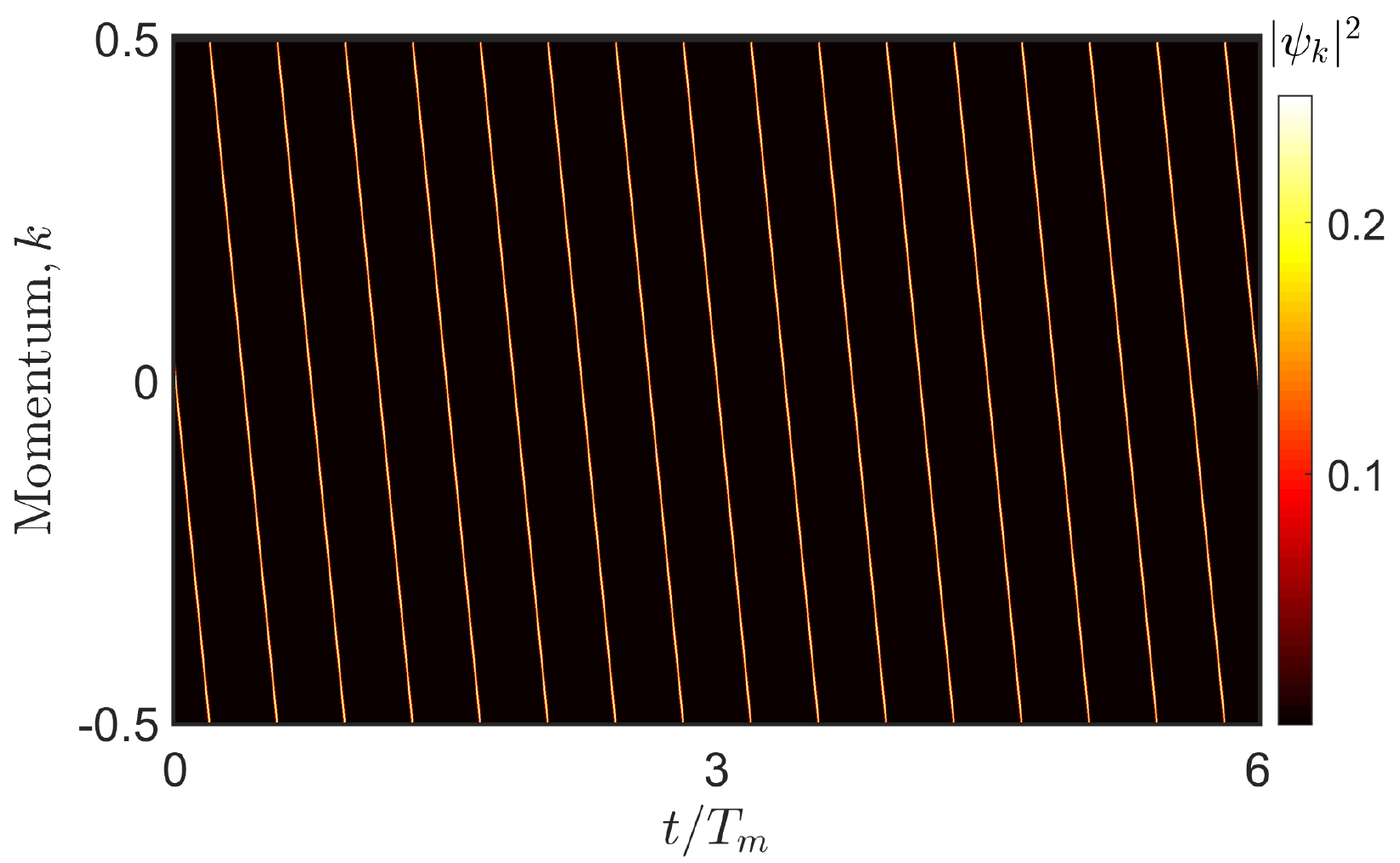}
	\caption{Time evolution of density distribution in momentum space. The parameters are chosen as $J=-1$, $\Delta_0=2$, $\delta_0=0.8$, $\omega=0.03$, $\phi_0=0$, $\omega_F/\omega=4/3$, $d=2$, $\sigma=15$, $j_0=101$ and $k_0=0$.}\label{FigMomentum}
\end{figure}

We are also interested in the density distribution in the quasi-momentum space, which is given by
\begin{eqnarray}
|\psi_k|^2=|\alpha_{o,k}|^2+|\alpha_{e,k}|^2,
\end{eqnarray}
with
\begin{eqnarray}
\alpha_{e,k} &=& \frac{1}{\sqrt{L}}\sum\limits_{j=1}^{L} {{e^{-ik2j}}\psi_j}, \nonumber \\
\alpha_{o,k}  &=& \frac{1}{\sqrt{L}}\sum\limits_{j=1}^{L} {{e^{-ik(2j - 1)}}\psi_j},
\end{eqnarray}
where $L=150$ is the total number of unit cells.  Fig.~\ref{FigMomentum} shows the time evolution of density distribution $|\psi_k|^2$ in the momentum space. In the whole time duration shown, the momentum distribution of the system maintains the same. The momentum is actually linearly swept down according to $K=k_0-\omega_F t$ and jumps to $0.5\pi$ when it reaches the boundary of Brillouin zone at $-0.5\pi$, which is consistent with the result in Sec.~\ref{MomentumSpace}.

\subsection{Time evolution of Guassian wave-packets} \label{EvolutionGuassian}

\begin{figure}[!t]
	\center
	\includegraphics[width=0.48 \textwidth]{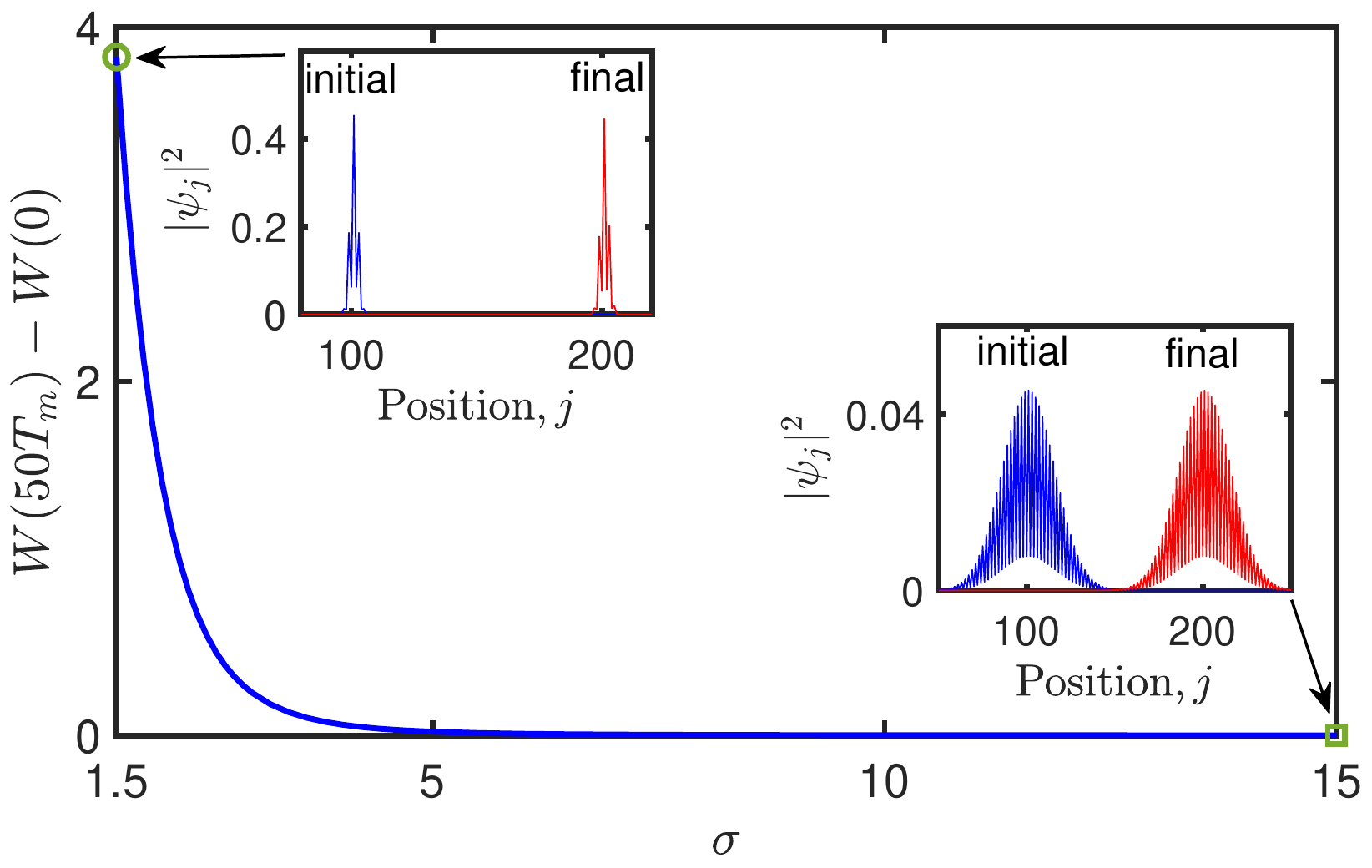}
	\caption{Difference in spatial width between final and initial states as a function of the initial width.
		Insets: Comparison between initial ($t=0$) and final ($50T_m$) states.
		The initial widths are chosen as $\sigma=1.5$ and $\sigma=15$ for the left and right insets, respectively.
		Other parameters are chosen as $J=-1$, $\Delta_0=2$, $\delta_0=0.8$, $\omega=0.03$, $\phi_0=0$, $\omega_F=4\omega$, $d=2$, $j_0=101$ and $k_0=0$.} \label{CompareInitialFinal}
\end{figure}

Even when the external force is integer multiples of the driving frequency ($\omega_F=n\omega$), a Gaussian wavepacket still undergoes quantized drifting Bloch oscillations due to the nontrivial Berry curvature.  This physics is certainly different from the early-observed possibility of resonance-induced expansion~\cite{Tarrallo}.
In Fig.~\ref{CompareInitialFinal}, we show difference in wavepacket width between the final and initial states for $\omega_F/\omega=4$.

A wider Guassian wavepacket in real space corresponds to a narrower Guassian wavepacket in momentum space, which is hence closer to a Bloch  state of a single quasi-momentum value.
We compare the density distribution of initial and final states in the insets of Fig.~\ref{CompareInitialFinal}.
For wide wavepackets, as they are more similar to a Bloch state, their density profiles are almost kept unchanged in their time evolution.
For narrow wavepackets, as they involve several  momentum states and different momentum states accumulate different phases~\cite{,Ke2017,Hu2019}, their shapes are slightly changed.

\begin{figure*}[!htp]
	\center
	\includegraphics[width=0.95 \textwidth]{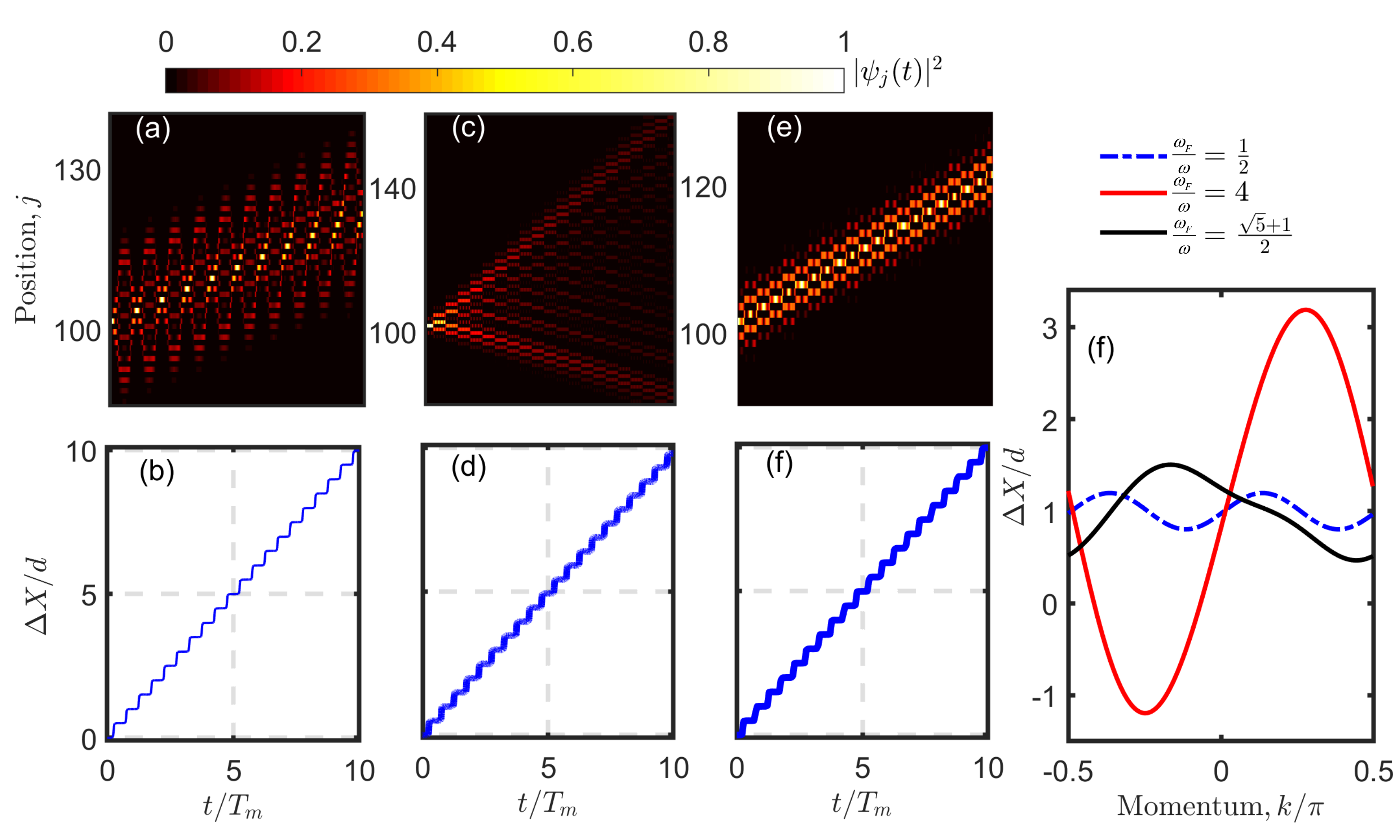}
	\caption{Topological pumping of an initial Wannier states. Panels (a), (c) and (e): Time evolution of density distribution for $\omega_F/\omega=1/2,\ 4,\ (\sqrt{5}+1)/2$, respectively. Panels (b), (d) and (f): Displacement as a function of time  for $\omega_F/\omega=1/2,\ 4,\ (\sqrt{5}+1)/2$, respectively. (g) The semi-classical displacement in one pumping cycle as a function of momentum for different $\omega_F/\omega$. The other parameters are chosen as $\Delta_0=20$, $J=-1$, $\delta_0=0.8$, $\omega=0.02$, $\phi_0=0$.
	}\label{FigWannierF4}
\end{figure*}

\section{Topological pumping of Wannier states} \label{PumpingWannier}
\subsection{General theory}
We consider the time-evolution of an initial Wannier state in the lower-energy band.
In contrast to Gaussian-like state localized at certain momentum, the initial Wannier state is an equal superposition of all the Bloch states in the lower-energy band with different momenta,
\begin{equation}
|w_1(R,0)\rangle=\frac{1}{\sqrt{L}}\sum\limits_k e^{-ikR}|u_{1}(k,0)\rangle,
\end{equation}
where $R$ denotes the location of the Wannier state.
When the system is adiabatically and periodically modulated, the displacement for the Wannier state in one pumping cycle is simply the average of the displacement for the Bloch states with different momenta,
\begin{equation}
\Delta X(T_m) =\frac{d}{2\pi}\int_{-\pi/d}^{\pi/d}\int_{0}^{T_m} v_g(k,t) dt dk, \label{DeltaXTm}
\end{equation}
where  $d$ as the period of the superlattice is equal to $2$ in our case.
The term due to the dispersion of the energy band is exactly zero, i.e.,
\begin{equation}
\frac{d}{2\pi}\int_{-\pi/d}^{\pi/d} \int_{0}^{T_m} \frac{\partial \varepsilon(k,t) }{\hbar \partial k} dt dk=0,
\end{equation}
regardless of the external force.
It means that the mean position shift in one pumping cycle is only related to the Chern number defined in the parameter space $(-\pi/2 \le k\le \pi/2,\ 0\le t\le T_m)$,
\begin{equation}\label{Chern}
C_n=\frac{1}{2\pi}\int_{-\pi/2}^{\pi/2}dk\int_{0}^{T_m}dt \mathcal F_n(k,t).
\end{equation}
Consequently, the displacement in one pumping cycle is given by
\begin{equation}
\Delta X(T_m)=C_n d,
\end{equation}
which is essentially the polarization theory~\cite{King1993}.

\subsection{Dynamics of Wannier states for different $\omega_F/\omega$}
It is known that the Bloch oscillations behave as breathing modes for an initial state localized at a single site~\cite{Hartmann2004}.
Such an initial state is approximately a Wannier state if the energy bands are flat.
The initial state is a single atom at an odd site (i.e. the $101$th site), which is approximately a Wannier state of the lower-energy band.
In Fig.~\ref{FigWannierF4}, we show the evolution of density distribution $|\psi_j(t)|^2$, the displacement $\Delta X(t)$ via quantum state evolution and the displacement as a function of momentum which is obtained via semi-classical expression [Eq.(6) in the main text].
In the numerical calculation, we choose $\Delta_0=20$ to make the energy bands flat. The other parameters are chosen as $J=-1$, $\delta_0=0.8$, $\omega=0.02$, $\phi_0=0$, $\omega_F=\omega/2$ for Fig.~\ref{FigWannierF4}(a) and \ref{FigWannierF4}(b), $\omega_F/\omega=4 $ for Fig.~\ref{FigWannierF4}(c) and \ref{FigWannierF4}(d), $\omega_F/\omega=(\sqrt{5}+1)/2$ for Fig.~\ref{FigWannierF4}(e) and \ref{FigWannierF4}(f).
In the case of $\omega_F=\omega/2$, it is clear that the wavepacket expands and shrinks periodically; see  Fig.~\ref{FigWannierF4}(a).
At the nodes of multiples of modulation period, the wavepacket is re-localized at a single site, but its mean position is shifted by a unit cell per pumping cycle.
In the absence of a tilt, the wavepacket is dispersive due to the curved energy band~\cite{ke2016topological}.
Here, the dispersion at the nodes is suppressed, because the group velocities of the momentum states have small fluctuations and hence their displacement, see the blue dashed-dot line in Fig.~\ref{FigWannierF4}(g).
Compared to the oscillation width of the time evolving  wave-packets, the quantized displacement is small but it is clearly found in the displacement; see Fig.~\ref{FigWannierF4}(b).
Note that $\omega_F=\omega/2$ is chosen for the coincidence between the period of Bloch oscillations and th period of pumping cycle.
In this case, the period of Bloch oscillations is determined by the time $T=\pi/\omega_F=2\pi/\omega$ to sweep the first Brillouin zone.
Besides, the dispersion of wavepacket still exists in the long time evolution due to the dispersion of energy band and Berry curvature.

In the case of $\omega_F/ \omega=4$,  instead of the breathing modes, the system becomes diffusive like a bullet mode while the mean position shift per period remains quantized; see Fig.~\ref{FigWannierF4}(c) and \ref{FigWannierF4}(d).
The diffusion becomes faster than the previous case, because the mean position shift has larger fluctuation as the momentum changes;  see the red solid line in Fig.~\ref{FigWannierF4}(g).

For completeness, we also consider the dynamics in the irrational case where $\omega_F/\omega=(\sqrt{5}+1)/2$; see Fig.~\ref{FigWannierF4}(e) and (f).
Compared with the irrational case in Sec.~\ref{Gaussian}, the time evolution of the density distribution has no well-defined period and behaves as quasi-periodic breathing modes.
As the strength of the tilt increases, the width of the breathing mode becomes smaller.
Nevertheless, the displacement is quantized per each pumping cycle.
The fluctuation of mean position shift with momentum is presented as the black solid line in Fig.~\ref{FigWannierF4}(g).
In all the above cases, although the details of the dynamics are quite different for different ratios between the Bloch oscillation frequency and the modulation frequency, the quantized displacement in one pumping cycle maintains the same, which is consistent with the theory in Sec.~\ref{Gaussian}.

\section{Experimental feasibility} \label{Experiment}

In a time-modulated superlattice subjected to an external force, the motion of ultracold atoms (e.g. Rubidium atoms) is governed by a continuous Hamiltonian,
\begin{equation}
\hat H_0 =-\frac{\hbar^2}{2m} \frac{{\partial ^2}}{\partial x^2} - {V_s}{\cos ^2}\left( {\frac{\pi }{a}x} \right) - {V_l}{\cos ^2}\left( {\frac{\pi }{{2a}}x - \frac{\phi} {2}} \right)
+\frac{F}{a}x. \nonumber \\
\end{equation}
Here, $m$ is the mass of the atom. $V_s$ and $V_l$ are the strengths of optical lattices with lattice spacing $a$ and $2a$, respectively.  $a$ is half of the wavelength for the short lattice, and the spacing of a unit cell is $d=2a$.
$\phi(t)=\omega t+\phi_0$ is the linear-shifting phase between the short and long lattices.
Under tight-binding approximation, we can obtain the Hamiltonian~\eqref{Ham} in the main text.
We make a dimensionless transformation, $x'= x/a$ and $t'={E_r} t/{\hbar}$, and express the Hamiltonian in the unit of $E_r=\frac{\pi^2\hbar^2}{2ma^2}$,
\begin{eqnarray}
\hat H_{0}' &=&  - \frac{1}{\pi^2}\frac{{{\partial ^2}}}{{{\partial}x'^2}} - \frac{{{V_s}}}{2E_r}\cos \left( {2 \pi x'} \right) \nonumber\\
&-& \frac{{{V_l}}}{2E_r}\cos \left( {{\pi x'} - \frac{\hbar \omega}{E_r} t'-\phi_0  } \right)+\frac{F}{E_r} x'.
\label{origin}
\end{eqnarray}

\begin{figure}[htp]
	\center
	\includegraphics[width=0.5 \textwidth]{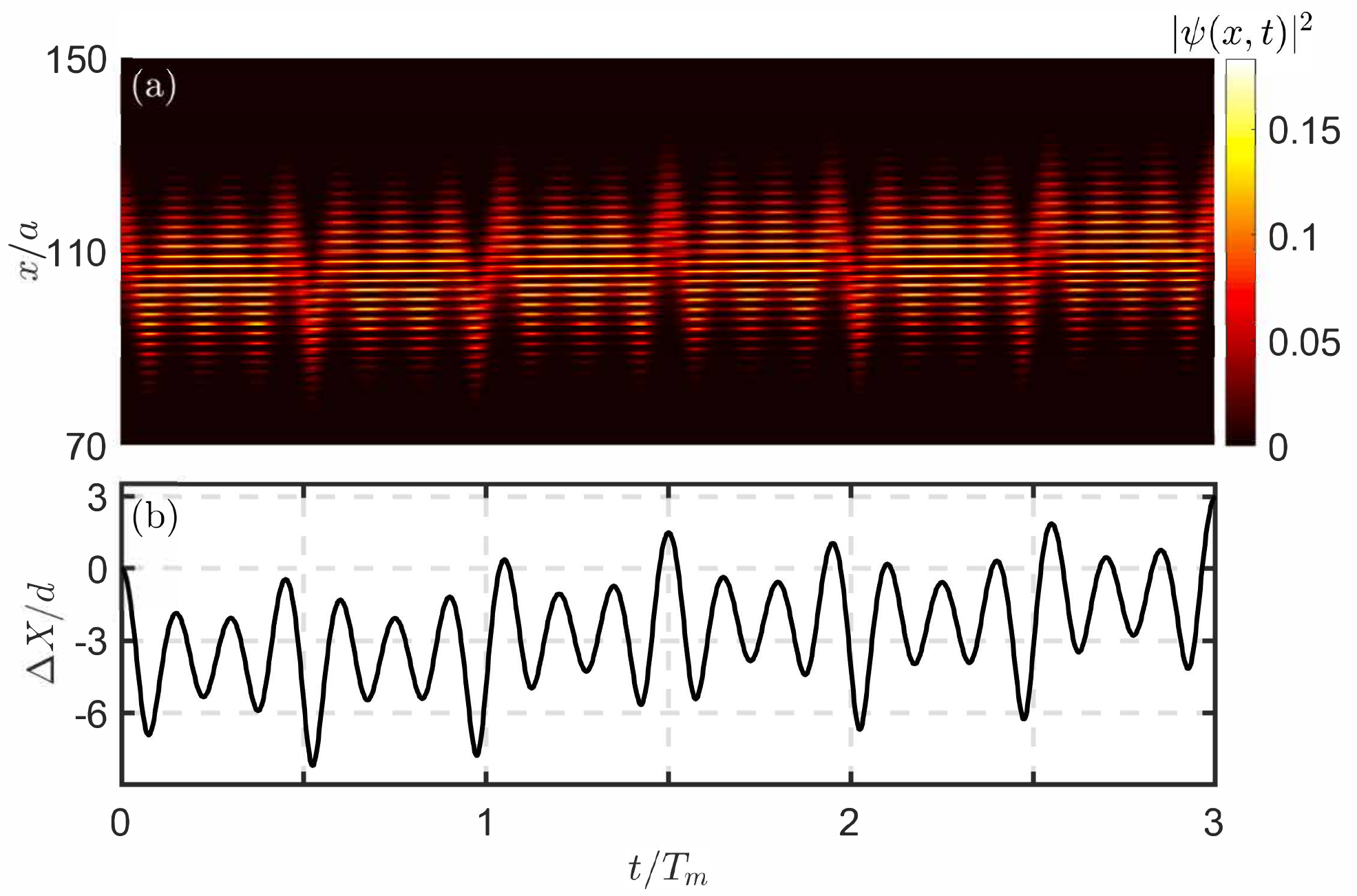}
	\caption{Dynamics of a Gaussian wavepacket governed by the original Hamiltonian. (a) Probability distribution and (b) displacement as functions of time in an overall period. Numerical calculations are performed with $V_s=2 E_r$, $V_l=E_r$, $\hbar \omega=0.002 E_r$, $F/(\hbar \omega)=10/3$, $\phi_0=\pi/2$, $x_0/a=112$, and $\gamma=10^{-5}E_r$.
	}\label{OriDynamics}
\end{figure}

The initial state is prepared as the ground state in a superlattice subjected to an additional harmonic trap instead of a tilted field, $V_{trap}=\frac{1}{2} \frac{\gamma}{a^2} (x-x_0)^2$ with the strength $\gamma$ and center position $x_0$, i.e., $V_{trap}'=\frac{1}{2} \frac{\gamma}{E_r} (x'-x_0')^2$ in the unit of $E_r$.
In the numerical calculations, the ground state is obtained by the method of imaginary time evolution.
Once the initial state has been prepared, we turn off the harmonic trap and turn on the tilted field and leave the Gaussian wavepacket evolve under the Schr\"{o}dinger equation,
$i\frac{\partial}{\partial t'}|\psi(t')\rangle=\hat H_0' |\psi(t')\rangle$, by a spectral method~\cite{Feit1982}.
The parameters are chosen as $V_s=2 E_r$, $V_l=E_r$, $\hbar \omega=0.002 E_r$, $F/(\hbar \omega)=10/3$, $\phi_0=\pi/2$, $x_0/a=112$, and $\gamma=10^{-5}E_r$.
In Fig.~\ref{OriDynamics}, we show the probability distribution and the mean position shift as functions of time.
The displacement in an overall period is $0.9995$ times $qd$, almost a quantized value as predicted in the main text.
Due to the topological nature of this experimental proposal, one does not need to fine-tune the strengths of the short and long lattices and the initial phase difference between the short and long lattices.
The modulated frequency can also be precisely controlled to satisfy its rational relation with the tilted field.
We believe that our new pumping scheme can be readily realized with the state-of-art techniques of cold atomic systems.

\bibliography{DriftBloch}

\end{document}